\documentclass[fleqn,usenatbib]{rasti}

\usepackage[dvipsnames]{xcolor}
\usepackage{newtxtext,newtxmath}

\usepackage[T1]{fontenc}

\DeclareRobustCommand{\VAN}[3]{#2}
\let\VANthebibliography\thebibliography
\def\thebibliography{\DeclareRobustCommand{\VAN}[3]{##3}\VANthebibliography}

\usepackage{graphicx}	
\usepackage{amsmath}	
\usepackage{caption}
\usepackage{makecell}
\usepackage{subcaption}
\usepackage{multicol}
\usepackage{multirow}
\usepackage{tabularx}
\usepackage{units}
\UseRawInputEncoding
\usepackage{lineno}

\title[Polychromatic PWFS with MKID technology]{Polychromatic pyramid wavefront sensor with MKID technology for high contrast imaging}

\author[A. Magniez et al.]{
Aurélie Magniez$^{1}$\thanks{E-mail: aurelie.magniez@durham.ac.uk},
Charlotte Z. Bond$^{2}$,
Peter Wizinowich$^{1, 3}$,
Tim Morris$^{1}$,
Kieran O'Brien$^{1}$.
\\
$^{1}$Durham University, Department of Physics, South Road, Durham, DH1 3LE, UK\\
$^{2}$UK Astronomy Technology Centre, Royal Observatory, Edinburgh, EH9 3HJ, UK\\
$^{3}$W. M. Keck Observatory, Kamuela, HI 96743, USA\\
}

\date{Accepted XXX. Received YYY; in original form ZZZ}

\pubyear{\the\year{}}

\begin{document}
\label{firstpage}
\pagerange{\pageref{firstpage}--\pageref{lastpage}}
\maketitle

\begin{abstract}
 
 The high sensitivity of the pyramid wavefront sensor has made it  the preferred sensor in high contrast adaptive optics systems. Future higher contrast systems, like the Extremely Large Telescope's Planetary Camera System, will require higher performance wavefront sensing. A further performance improvement could be achieved with a polychromatic pyramid wavefront sensor by using additional information over a broader wavelength range. The development of such systems is becoming more feasible with the emergence of new detector technologies such as Microwave Kinetic Inductance Detector arrays. These are arrays of superconductor detectors that give a position, arrival time and measure of the energy for each incident photon. This paper introduces the polychromatic pyramid wavefront sensor concept by defining the technologies and techniques employed and their requirements. A method is developed to track the optical gains, taking advantage of the additional wavelength information, and used to compensate for optical gains within an optimised reconstructor to minimise noise propagation. An overview of expected performance improvement, using end-to-end simulations, is provided using the Keck II adaptive optics system as a reference design. The polychromatic pyramid wavefront sensor was shown to increase the limiting magnitude by 1 to 2 magnitudes, and the contrast by factors of 1.5 to 4, versus single band pyramid wavefront sensors, by sensing over a wavelength range approximately five to ten times broader (800–1800 nm) compared to Z band (152 nm wide) and H band (300 nm wide). Practical design and implementation issues have also been considered.
\end{abstract}




\section{Introduction}

Adaptive Optics (AO), which corrects for the wavefront distortions introduced by atmospheric turbulence, is a key technology that significantly improves the observation capabilities of ground-based telescopes. As we move towards the era of Extremely Large Telescopes (ELTs), new challenges emerge, particularly for high-contrast observations, leading to research for new technologies and techniques. Future high contrast instruments, such as the ELT's Planetary Spectrograph Camera (PCS; \cite{kasper_pcs_2021}), will require unprecedented angular resolution, calling for large-scale extreme AO systems and improvements in wavefront sensing. The pyramid wavefront sensor (PWFS; \cite{ragazzoni_pupil_1996}) achieves high sensitivity and is the wavefront sensor (WFS) of choice for future extreme AO systems. A crucial element of the PWFS is its detector which needs to satisfy requirements of low noise (<1e-), high speed (kHz frame rates) and sufficient spatial resolution for the wavefront measurement (10s to 100s of pixels across the pupil). While existing technologies satisfy these requirements for current instruments, research into more capable detectors continues for the next generation of instruments. This raises the question of how performance can be enhanced by integrating advanced detector technologies that provide more information.

In this context, we propose the use of an energy sensitive detector: a Microwave Kinetic Inductance Detector (MKID, \cite{day_broadband_2003}) array, that provides for each photon a measure of its energy, an arrival time and a position. By bringing this new information, an MKID array can extend the performance of a WFS compared to the capabilities provided by traditional semiconductor detectors. While MKID technology is still under development, it shows considerable potential for wavefront sensing. Its energy sensitivity allows for simultaneous measurements at multiple colour bands, increasing the photon count for wavefront measurements. Additionally, the temporal and chromatic information it provides can help correct for wavefront measurement errors, increasing the accuracy of the measurements.

Preliminary on-sky results with a polychromatic PWFS were first reported by \cite{magniez_mkid_2022, magniez_polychromatic_2024}. This paper extends this earlier work by fully defining the concept and quantifying its expected performance.  Complementary work by \cite{darcis_adding_2025} evaluated a multi-wavelength Zernike wavefront sensor that uses additional photons to extend the dynamic range beyond that of a scalar Zernike wavefront sensor.

Pushing this research further, the present paper introduces the concept of a polychromatic PWFS. Firstly, we describe how an MKID array designed for extreme AO system could be used, its requirements, and the architecture of a polychromatic PWFS (section \ref{sec:polychromatic_pwfs}). We then detail the behaviour of a PWFS at different wavelengths (section \ref{sec:polychromatic_behaviour}). We propose a method to quantify the non-linearities introduced by phase residuals (optical gains) and demonstrate how they can be compensated using only the polychromatic reconstruction of the wavefront (section \ref{sec:OG_tracking}). We simulate the integration of a polychromatic wavefront sensor into an existing system, using the Keck II Telescope AO system as a case study (section \ref{sec:simulation_study}). Finally, we discuss the practical considerations for deploying such a system on-sky (section \ref{sec:practicalities}).

\section{A Polychromatic PWFS}
\label{sec:polychromatic_pwfs}

\subsection {Microwave Kinetic Inductance Detectors for adaptive optics}

\begin{figure}
\centering\includegraphics[width=0.7\linewidth,trim={0cm .8cm 0cm 0cm}]{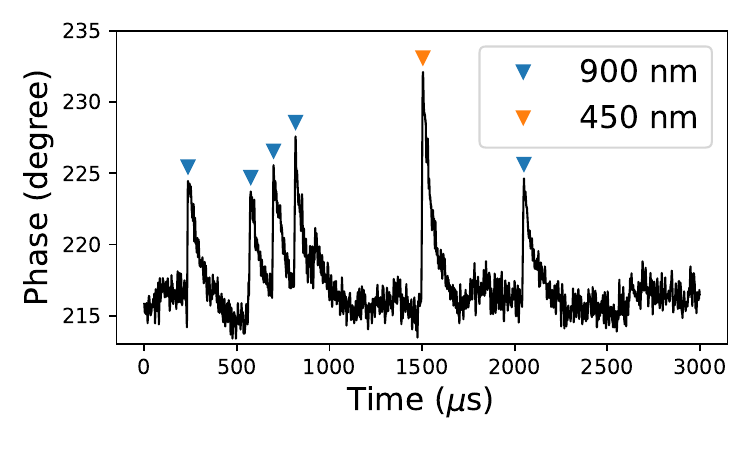}
\caption{An example of a MKID output data stream showing the characteristic resonator phase change against time for detection of six different photons over 3 milliseconds at wavelengths of 450~nm and 900~nm. }
\label{fig:pulse}
\end{figure}

An MKID is a superconducting detector with the capability to provide a measurement of the energy and arrival time of an incident photon. As described in \cite{mazin_superconducting_2012}, each MKID pixel is an inductor-capacitor resonator into which a probe microwave signal is sent. In the optical/NIR regime, the energy of an incident photon is sufficient to break thousands of Cooper pairs, which leads to a change in inductance in the resonator through the kinetic inductance effect \cite{day_broadband_2003}. This can be measured as a change in phase of the microwave signal in the form of a fast-rise, exponential decay pulse, as seen in Fig. \ref{fig:pulse}. The pulse amplitude is proportional to the energy of the photon. Moreover, because the energy deposited by an optical/NIR photon is many times the threshold set for detection, the readout does not introduce additional false counts, and MKIDs are therefore often described as having zero readout noise. To decrease the sensitivity to longer wavelength photons, optical filters are used to supress the out of band flux (e.g. from the thermal infrared part of the spectrum). In this work, we consider a sample range of 800 to 1800 nm

Unlike other energy-sensitive detector technologies (see \cite{oconnor_energy-sensitive_2019} for a review) MKIDs can easily be multiplexed into large arrays \citep{mchugh_readout_2012}. MKID pixels are typically sampled at 1~MHz,  currently limited by the readout electronics,  providing a photon arrival time accuracy of $\sim$1~$\mu$s.  The output of the MKID is the measurement of the photon energy and the arrival time of each photon for each pixel in the array. From this, we can build a frame with greater flexibility in the reconstruction of an image, where one could, for example, impose a fixed signal-to-noise ratio (SNR) by dynamically adjusting the frame rate based on detected photons. The effective frame rate enabled by an MKID array exceeds that of existing detectors currently used for AO (as discussed below). Fig. \ref{fig:pulse} shows the output data stream when several photons are incident on the detector within a short space of time.  Alternative detector technologies, such as SPAD arrays \citep{carrier_photon--digital_2023}.  can achieve faster temporal response, but cannot provide energy sensitivity and do not currently scale to the number of pixels required for high-contrast AO wavefront sensing. 

The first MKID-based instrument for optical/IR astronomy was ARCONS (ARray Camera for Optical to Near-IR Spectrophotometry) at the Palomar Observatory in 2010 \citep{mazin_arcons_2013}. It contained a 2024-pixel MKID array, which had an energy resolution of $\sim$5 and operated in the range 400 to 1200~nm with a detector quantum efficiency (QE) of $\sim$0.2. Since this first demonstration, larger arrays have been made, leading to the 20,440 pixel array in the Subaru telescope's MKID Exoplanet Camera (MEC; \cite{walter_mkid_2020}). This array can effectively discern up to 9 wavebands, directly derived from the MKID energy bands, between 400 and 1700~nm. Further advances in energy resolution and QE have been demonstrated at the single-pixel level and are currently being incorporated into the designs for arrays \citep{zobrist_membraneless_2022, de_visser_phonon-trapping-enhanced_2021}. 

In high-flux conditions, the arrival of two or more photons on the same pixel leads to a phenomenon where a second photon arrives before the baseline has returned to its equilibrium state. This makes it more difficult to determine the pulse height and may even appear as a single higher energy photon if both photons arrive within the rise time of the pulse. This leads to a maximum flux rate requirement, which can be managed through a combination of instrument design (e.g. oversampling), array design (e.g. faster recombination timescales), or read-out design (e.g. more complex pulse detection algorithms).

Table \ref{tab:comparison_detectors} shows the AO-relevant characteristics for current EMCCD \citep{feautrier_characterization_2010} and Saphira \citep{feautrier_last_2022} detectors, compared to the MEC array \citep{walter_mkid_2020}. We also show potential characteristics of a future MKID array optimised for a high-contrast ELT-like instrument. For our example, we assume a pupil sampling similar to current extreme AO systems \citep{sauvage_saxo_2016} (40 pixels across 8 m) scaled to the ELT 39 m pupil resulting in $\sim$ 200 pixels across the pupil diameter. As MKID pixel geometries are not limited to square arrays, we have defined the number of pixels as the minimum needed to sample the four illuminated circular pupils of the PWFS. Unlike existing WFS detectors, MKID arrays can provide high QE across the full visible and near-infrared spectral range simultaneously.  Laboratory measurements show >80\% intrinsic QE between 400 and 1550 nm \citep{kouwenhoven_model_2022} for the films used in MKID devices, but a practical measurement of an antireflection-coated MKID remains to be made.  We note that when using an EMCCD in high-gain without photon-counting, as is typically the case for astronomical AO, the electron multiplication causes an excess of noise equivalent to a 50\% reduction in QE \citep{robbins_noise_2003}. That is not the case for the photon-counting MKID.

\begin{table}
    \centering
    \caption{Comparison between the EMCCD and Saphira detectors commonly used in astronomical AO systems, the current MKID used in MEC and the expected performance for an ELT-scale instrument PWFS MKID.}
    \begin{tabular}{l c c c |c}
    \hline
      & \textbf{Saphira} & \textbf{EMCCD} & \textbf{MEC} & \textbf{MKID}\\
      
    \hline
    Pixel count & 81920 & 57600 & 20440  & >126000\\
    Read noise (e$^{-}$) & 0.6 & 0.5 & 0 & 0 \\
    Sampling (kHz)  & 3.5 & 2 & 1000 & 1000 \\
    Pixel size ($\mu$m)  & 24 & 24 & 150 & 150\\
    Quantum Efficiency& 0.8 & 0.9 & 0.3 & 0.8 \\
    Wavelength range& IR & Vis & IR+Vis & IR+Vis \\
    Colour bands & 1 & 1 & 9 & >9\\
    \hline
    \end{tabular}

    \label{tab:comparison_detectors}
\end{table}

\subsection{Polychromatic MKID-based PWFS}
\label{sec:PolyPWFS}

A PWFS typically consists of a four-sided pyramid prism, re-imaging optics and a detector. Light from the telescope is focused onto the tip of the pyramid, which splits the incoming light onto four paths. A re-imaging lens conjugates the pupil plane with the detector, producing four images of the telescope pupil \citep{ragazzoni_pupil_1996}. When the incoming wavefront is perturbed, the relative intensity distribution between the four pupil images is modified, enabling a measurement of the wavefront. The PWFS has been widely adopted for the next generation of high-contrast instruments due to its high sensitivity. By using an MKID array for the PWFS detector the sensor can measure the wavefront at multiple wavelengths. In this section we outline the potential advantages of such a polychromatic PWFS.

\textit{More photons for wavefront sensing.} The use of a polychromatic PWFS would allow for the use of a wide range of wavelengths simultaneously, potentially increasing the number of photons available for the wavefront measurement and extending the limiting magnitude. An example of the increased photons and improved limiting magnitude is provided in section \ref{sec:ao_system_keck}. 

\textit{Improved wavelength calibration.} A PWFS operating over a single defined waveband has no knowledge of the wavelength of the incident photons, beyond the bandpass of any dichroics or filters placed upstream of the PWFS. Whilst the response is calibrated for a known wavelength, the weighted average wavelength during operation will depend on the colour of the guide star, with different star types therefore producing different responses. A polychromatic PWFS can operate over a wider range of wavelengths, using the wavelength information to avoid such chromatic errors which can bias the reconstruction. 

\textit{Increased linearity and/or sensitivity.} AO performance is a trade off between sensitivity (favouring shorter wavelengths) and robustness to non-linearities (favouring longer wavelengths) \citep{correia_performance_2020}. The use of a polychromatic PWFS, where a wide range of wavelengths are used, has the potential to optimise the reconstruction for different conditions, taking advantage of the increased linear range of longer wavelengths to optimise the peak performance for bright guide stars, whilst utilising the higher sensitivity at shorter wavelengths to target fainter stars. This trade off is discussed in section \ref{sec:linearity}. 

\begin{figure}
\centering\includegraphics[width=0.85\linewidth]{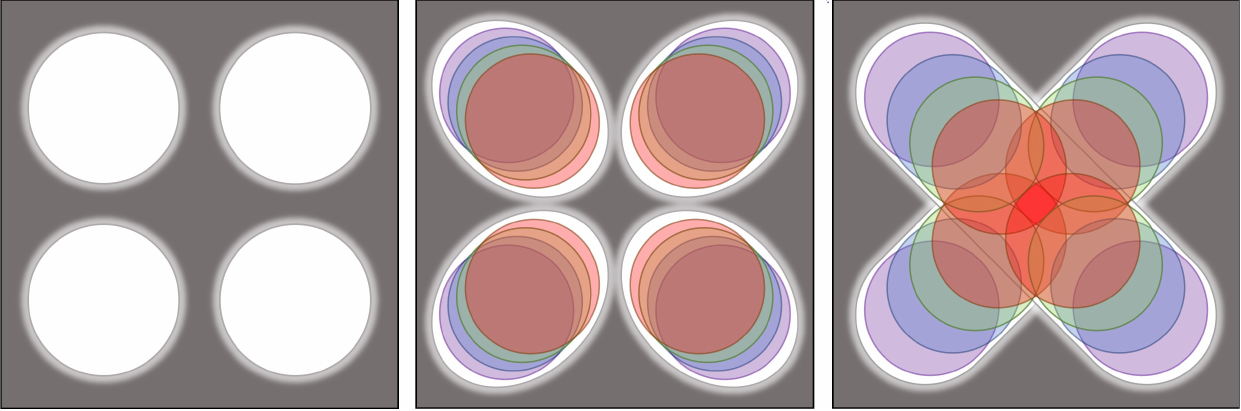}
\caption{Illustration of different intensities read by a PWFS detector. From left to right: non-energy sensitive detector detecting a narrow waveband, MKID detecting multiple wavebands, MKID detecting multiple wavebands with an exaggerated dispersion and a 'flattened pyramid' case for the central pupil images.}
\label{fig:pupil dispersion}
\end{figure}

\textit{Reduced pupil constraints.} Wavelength information can also be used to address the impact of chromatic aberrations and dispersion within the AO system. The pyramid itself causes dispersion at the pupil plane as illustrated in Fig.~\ref{fig:pupil dispersion} which in a non-polychromatic PWFS results in smearing of the wavefront signal across the pupil and a reduction in sensitivity at higher spatial frequencies. This is typically managed by the use of double pyramids comprised  of two glasses with complementary refractive indices chosen to minimise the dispersion over the chosen wavelength range ~\citep{schwartz_design_2020}, in combination with achromatic pupil reimaging optics. 
However, even these are limited to a specific range of wavelengths. A polychromatic PWFS would allow separation of the chromatically dispersed pupils~\citep{magniez_mkid_2022} providing an expansion of the overall waveband of the sensor and reducing the dispersion requirements, allowing for alternative, simpler PWFS optical implementations to be considered \citep{lardiere_double-pyramid_2017}. 

\textit{Improved spatial resolution or sensitivity.} We may also be able to take advantage of the diversity offered by the chromatic dispersion of the pupils (Fig.~\ref{fig:pupil dispersion}) to explore the possibility of \textit{chromatic} super-resolution~\citep{oberti_super-resolution_2022} and combinations of different PWFS geometries such as the flattened pyramid~\citep{fauvarque_general_2017}.

\textit{Atmospheric dispersion control.} Other wavelength-dependent aberrations, such as residual atmospheric dispersion, could be measured and potentially corrected (e.g. fed back to an atmospheric dispersion corrector, as discussed in section \ref{sec:ADC}). 

\textit{Segment phasing.} The energy sensitivity of the MKID may also allow the PWFS to better operate as a segmented mirror piston sensor~\citep{hedglen_lab_2022} and detect large amplitude phase steps due to pupil fragmentation at the telescope spiders. Like many phasing sensors that operate over a narrow wavelength range, the PWFS exhibits a $\pm \lambda/2$ ambiguity in measured piston between segments. An MKID array designed to operate across the visible and near-infrared range would break this ambiguity, providing a large capture range segment phasing sensor. 

\textit{Optical gain optimization.} The additional wavelength information provided by a polychromatic PWFS can be used to improve the correction by exploiting the chromatic nature of other common errors in the WFS measurement. One characteristic of the PWFS which could be addressed is its non-linear behaviour, particularly at shorter wavelengths. This non-linearity is quantified by the \emph{optical gain}, a spatial frequency dependent sensitivity loss dependent on operating conditions (i.e. atmospheric seeing) and varying with wavelength~\citep{magniez_polychromatic_2024}. A polychromatic PWFS could be used to estimate (and eventually compensate) the optical gains in real time by monitoring the ratio of the reconstruction at different wavelengths. This optical gain tracking is a key focus of this paper and will be presented in sections~\ref{sec:OG},~\ref{sec:OG_tracking},~\ref{sec:OG_tracking_perf},~\ref{sec:OG_tracking_noise} and~\ref{sec:perf-improve}.

\section{Polychromatic behaviour of the PWFS}
\label{sec:polychromatic_behaviour}

In this section, we consider the behaviour of a polychromatic PWFS, specifically the linear range, sensitivity, optical gain and noise propagation, all as a function of wavelength. We will then use the theoretical noise propagation to propose a method for optimising the wavefront reconstruction from polychromtic PWFS signals.

\subsection{Linearity and sensitivity}
\label{sec:linearity}

\begin{figure}
    \begin{subfigure}[t]{1\linewidth}
        \centering
        \includegraphics[width=0.3\linewidth]{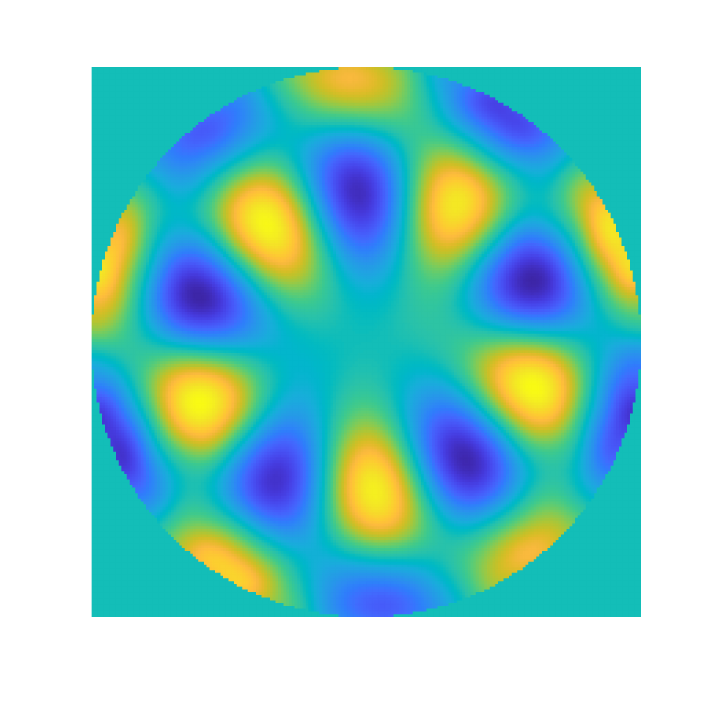}
        \caption{Example wavefront aberration (KL mode 30).}
    \end{subfigure}
    
    \vspace{3mm}
    
    \begin{subfigure}[t]{1\linewidth}
        \centering
        \includegraphics[width=0.98\linewidth]{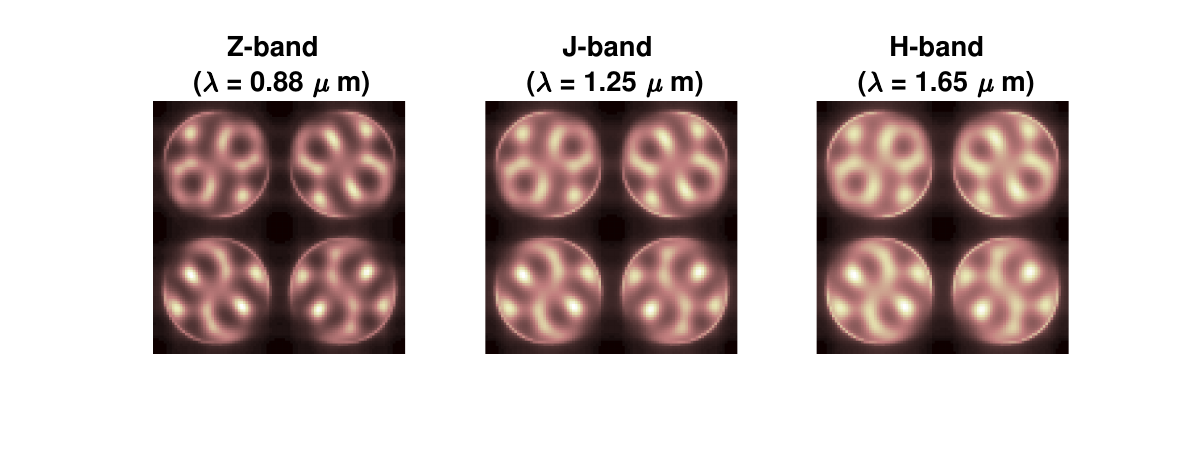}
        \caption{PWFS response at different wavelengths.}
    \end{subfigure}
    \caption{The PWFS response at different wavelengths to KL mode 30 with a wavefront rms of 100\,nm. In each case, the modulation is 3\,$\lambda/D$. The choice of wavelengths in this paper is discussed in section \ref{sec:ao_system_keck}.
    }
    \label{fig:PWFS_signals}
\end{figure}

\begin{figure}
    \centering
    \includegraphics[width=0.98\linewidth]{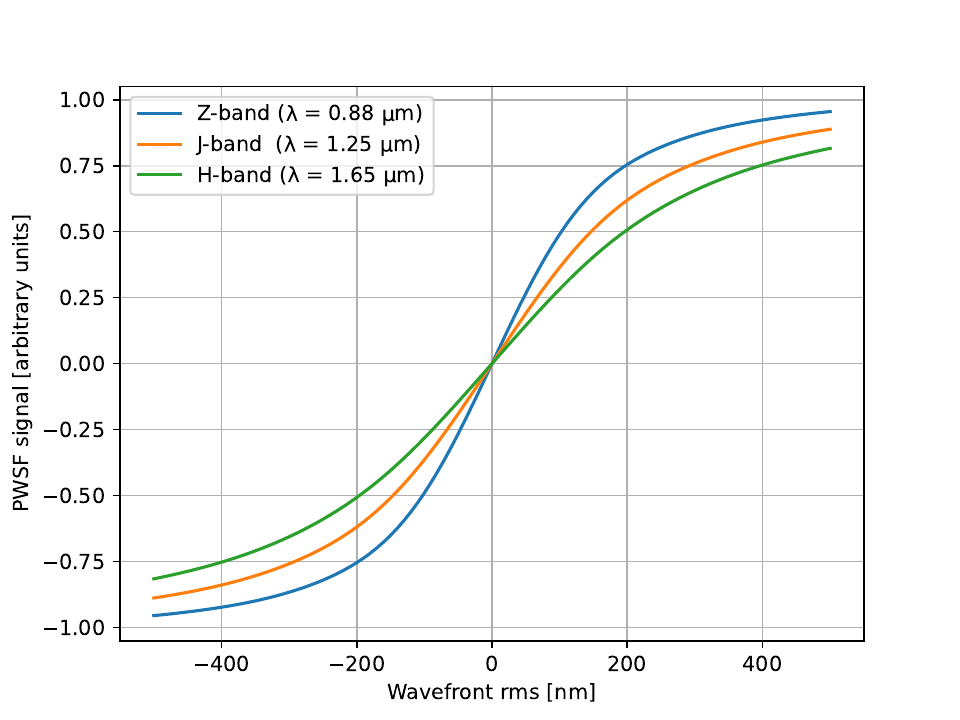}
    \caption{PWFS response to varying amplitude of an incoming wavefront aberration (KL mode 30). For each wavelength, the PWFS is operating with a modulation radius of 3\,$\lambda/D$.}
   
    \label{fig:linearity}
\end{figure}

The PWFS is a highly sensitive wavefront sensor. However, this sensitivity comes at the expense of a limited linear range.  Both the sensitivity and linearity of the sensor depend on the wavelength. If we first consider the response of a PWFS for a given small wavefront aberration (within the sensors linear range), we observe that the PWFS signal for shorter wavelengths is stronger than the signal at longer wavelengths. Fig.~\ref{fig:PWFS_signals} shows examples of the PWFS signal at different wavelengths for a given Karhunen-Loeve (KL) mode with a root mean square (rms) wavefront error of 100\,nm.  For longer wavelengths, the equivalent phase for a given wavefront is weaker and therefore induces smaller intensity variations on the PWFS. Looking at the response of the PWFS, as the amplitude of the wavefront aberration increases, we observe that the PWFS signal exhibits a linear response for a small range of amplitudes around zero, after which the signal begins to taper off, as illustrated in Fig.~\ref{fig:linearity}. The result is that a PWFS working at longer wavelengths has a greater linear range than a PWFS working at shorter wavelengths.

Whilst the linearity of the PWFS is best served by using longer wavelengths, the sensitivity benefits from operation at shorter wavelengths. Here we define sensitivity as the gradient of the PWFS response within the linear range. Fig.~\ref{fig:linearity} illustrates that the sensitivity increases as the wavelength decreases - essentially more photons are converted into a measurement of the wavefront at shorter wavelengths.

To increase the linear range, the PWFS is almost always modulated when operated on-sky. This involves modulating the point spread function (PSF) in a circle around the tip of the pyramid, typically using a tip-tilt mirror placed in an upstream pupil plane. The modulation period is then synchronised with the PWFS detector integration time. Modulating increases the effective size of the beam incident on the PWFS and results in a greater linear range for lower-order spatial modes, at the expense of some sensitivity.
The effect of modulating depends on the modulation radius compared to the size of the PSF. In the examples shown here we have used a modulation of 3\,$\lambda/D$ for each wavelength.
However, for a polychromatic PWFS the modulation will be a fixed angular amplitude and therefore, as a multiple of $\lambda/D$, will vary with wavelength. 

\subsection{Optical gain}
\label{sec:OG}

In the previous section, we considered the PWFS response for a single aberration, for an otherwise diffraction-limited PSF. When operating a PWFS on-sky, the PSF on the tip of the PWFS differs from the diffraction-limited PSF due to the AO phase residuals. The effect of these residuals is to broaden the PSF, resulting in a decrease in sensitivity. The extent of the PSF broadening will depend on atmospheric conditions (seeing) and AO performance. The loss in sensitivity between the calibration state (typically a diffraction limited PSF) and the operating state (an AO corrected PSF) is quantified by a parameter known as the optical gain. This is a modal vector with values between 0 and 1, where 1 indicates no loss in sensitivity from the calibration state. We define the modal coefficient of the wavefront reconstruction, $c_\lambda$, at a wavelength $\lambda$, for mode $\phi_i$, as a function of the optical gains ${g}_{opt}(\lambda, \phi_i)$, and the true wavefront reconstruction modal coefficient  ${c}_{true}(\phi_i)$:
\begin{equation}
    {c}_\lambda(\phi_i) = {g}_{opt}(\lambda, \phi_i) \times {c}_{true}(\phi_i).
    \label{eq:OG}
\end{equation}
 When operating with a bright guide star, high contrast AO systems are typically limited by fitting error, due to the limited number of corrected spatial modes \citep{guyon_limits_2005, correia_performance_2020}. In the following examples, we explore the impact of optical gains using a PSF limited only by fitting error.

To compute the optical gain we use a convolutional model of the PWFS~\citep{chambouleyron_pyramid_2020}. This model requires estimates of the PSF on the tip of the pyramid for both calibration (a diffraction limited PSF) and operation (the AO corrected PSF) to compute the response of the PWFS in both cases and subsequently the optical gains. In practice the PSF can reconstructed from AO telemetry or through direct measurement using a gain sensing camera\citep{chambouleyron_focal-plane-assisted_2021}, necessitating a portion of the wavefront sensor light to be diverted.

\begin{figure}
    \centering
    \includegraphics[width=.98\linewidth]{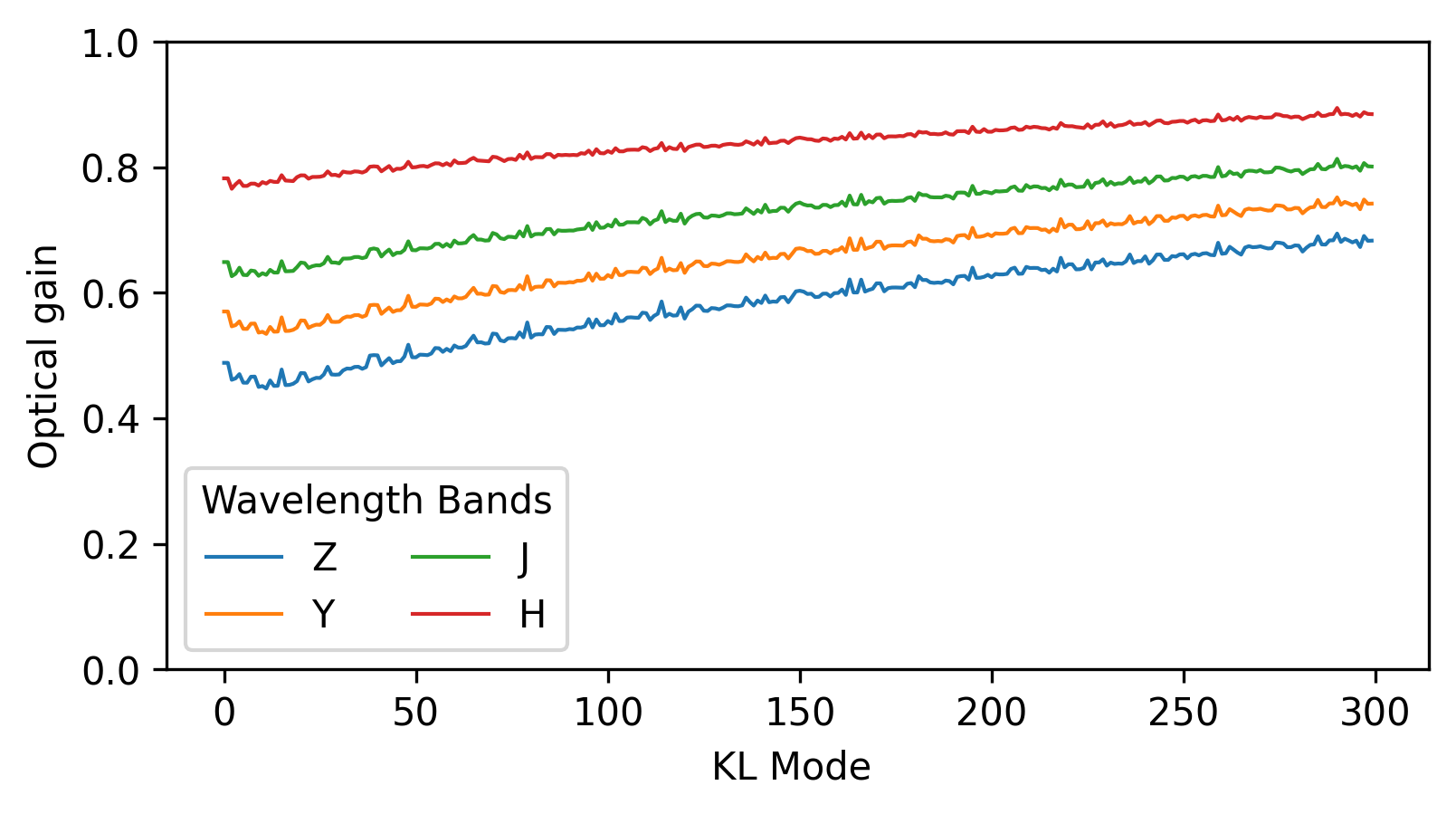}
    \caption{Optical gains computed with a fitting error PSF for $r_0 = 8$~cm at different wavelengths ($\lambda$ = 1.03 $\mu$m for Y-band) using the convolutional model. The parameters used for this simulation are given in Table \protect\ref{tab:params}.}.
    \label{fig:OG_at_various_wavelengths}
\end{figure}

Fig.~\ref{fig:OG_at_various_wavelengths} shows an example of the optical gains, computed for a fitting error PSF at different wavelengths for an $r_0$ of 8\,cm. The value of the optical gains vary with wavelength, with shorter wavelengths exhibiting a greater optical gain effect (smaller optical gains) as the PSF is further from the diffraction limit.  The optical gain is modally dependent and typically smaller for lower order modes, although the shape of the curve will depend on the morphology of the PSF. In the examples shown in Fig.~\ref{fig:OG_at_various_wavelengths} the optical gain for the shortest wavelength (Z-band) is smaller than 0.5 for the lower order modes. In other words the PWFS can underestimate the incoming wavefront by a factor of 2. 

\begin{figure}
    \centering
    \includegraphics[width = .98\linewidth]{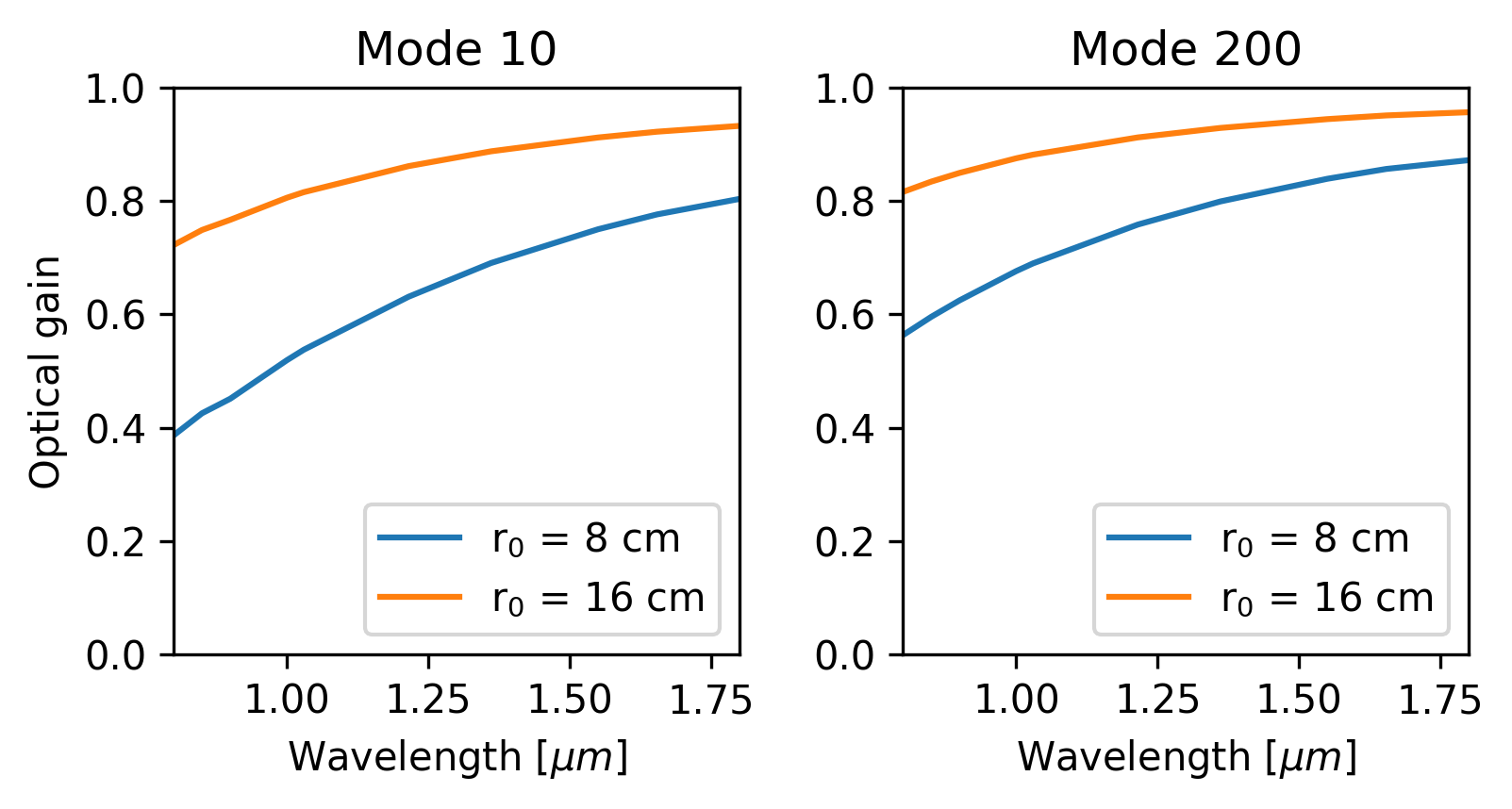}
    \caption{Chromatic variation of optical gains computed with a fitting error PSF at $r_0 = 8$~cm and $r_0 = 16$~cm using the convolutional model. The parameters used for this simulation are given in Table \protect\ref{tab:params}.}.
    \label{fig:Chromatic_OG_variation}
\end{figure}

Fig. \ref{fig:Chromatic_OG_variation} shows the variation with wavelength of the optical gain at  two  different atmospheric seeing conditions (i.e.  two values  of Fried parameter, $r_0$) for the KL modes 10 and 200. We can see that the chromatic variation is dependent on the strength of the atmospheric turbulence.  
 
Whilst the loss in sensitivity caused by AO residuals cannot be recovered, knowledge of the optical gain can be used to adjust the reconstruction of the wavefront to the on-sky response and maximise the AO performance. Typically, optical gain compensation is applied as the inverse of the optical gain estimate, either as an additional gain matrix or within the reconstructor itself.  In the example discussed above, where the Z-band PWFS has an optical gain $\sim0.5$ for the lower order modes, a compensation factor of $\sim 2$ would be applied. As the optical gain is modally dependent the compensation for the high order modes is smaller (corresponding to a larger optical gain). 

Optical gain compensation is particularly critical if the PWFS is working off null to compensate for non-common path aberrations. Several techniques currently exist to track the optical gains during operation~\citep{chambouleyron_pyramid_2020,deo_analyse_2019,esposito_-sky_2020}  However, these techniques typically require either additional hardware to monitor the PSF or the injection of disturbances into the AO control loop to directly measure the optical gain. In this paper, we propose that the use of a polychromatic PWFS can take advantage of the chromatic nature of the optical gains by inferring them from the ratios of the reconstructed wavefronts at different wavelengths. This idea is further expanded in section~\ref{sec:OG_tracking}.

\subsection{Noise propagation}
\label{sec:noiseProp}

As described in section~\ref{sec:linearity}, a PWFS working close to the diffraction limit is more sensitive when working at shorter wavelengths. This translates into a higher signal-to-noise ratio or alternatively a reduced noise propagation. Noise propagation refers to the error in the wavefront (or phase) measurement resulting from noise on the WFS. Here, we will consider the impact of photon noise and quantify the impact of noise at different wavelengths by computing the noise propagation due to photon noise. In this work, we therefore include only photon noise, as MKIDs do not exhibit additional intrinsic noise sources that generate false events; remaining effects (e.g. fabrication-related contributions to phase noise) merely raise the phase-detection threshold rather than limiting the long-wavelength cut-off or the achievable energy resolution.

The photon noise variance for an open loop PWFS measurement of phase $\phi_i$ is given by \cite{chambouleyron_modeling_2023}:
\begin{equation}
    \sigma^2({\phi_i}) = \frac{1}{s_{\mathrm{pn}}^2(\phi_i)\times N_{\mathrm{photons}}},
\label{eq:variance}
\end{equation}
where $N_{\mathrm{photons}}$ is the total number of photons per PWFS frame and $s_{\mathrm{pn}}(\phi_i)$ is the sensitivity with respect to photon noise for phase $\phi_i$:
\begin{equation}
    s_{\mathrm{pn}}(\phi_i) = \left|\left|\frac{\delta I(\phi_i)}{\sqrt{I_0}}\right|\right|_2.
\label{eq:sensitivity}
\end{equation}
$\delta I(\phi_i)$ is the normalised response of the PWFS to $\phi_i$ and $I_0$ is the normalised reference signal (typically the signal for a flat wavefront). 
To compute the noise propagation, for a given AO system, $\phi_i$ are the AO control modes and $\delta I (\phi_i)$ are the elements of the  interaction matrix (IM). To take into account the reduction in sensitivity, when operating on-sky, we multiply each element of the IM by the optical gain, $g_{opt}$, for that mode:
\begin{equation}
    \delta I (\phi_i) = g_{opt}(\phi_i) \mathrm{IM}(\phi_i).
    \label{eq:dI}
\end{equation}
\begin{figure}
    \begin{subfigure}[t]{1\linewidth}
        \centering
        \includegraphics[width=0.98\linewidth]{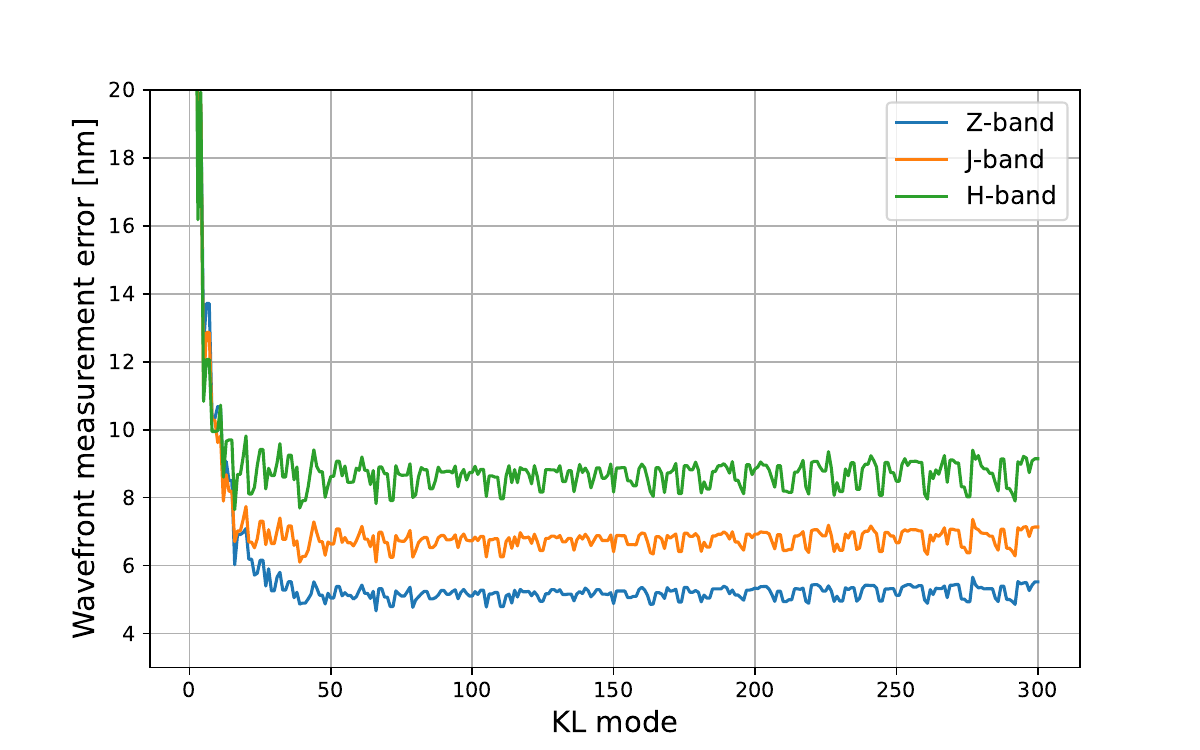}
        \caption{Wavefront measurement error for a diffraction limited PWFS operating at different wavelengths.}
        \label{fig:noiseVsKL_diffWaves}
    \end{subfigure}
    \begin{subfigure}[t]{1\linewidth}
        \centering
        \includegraphics[width=0.98\linewidth]{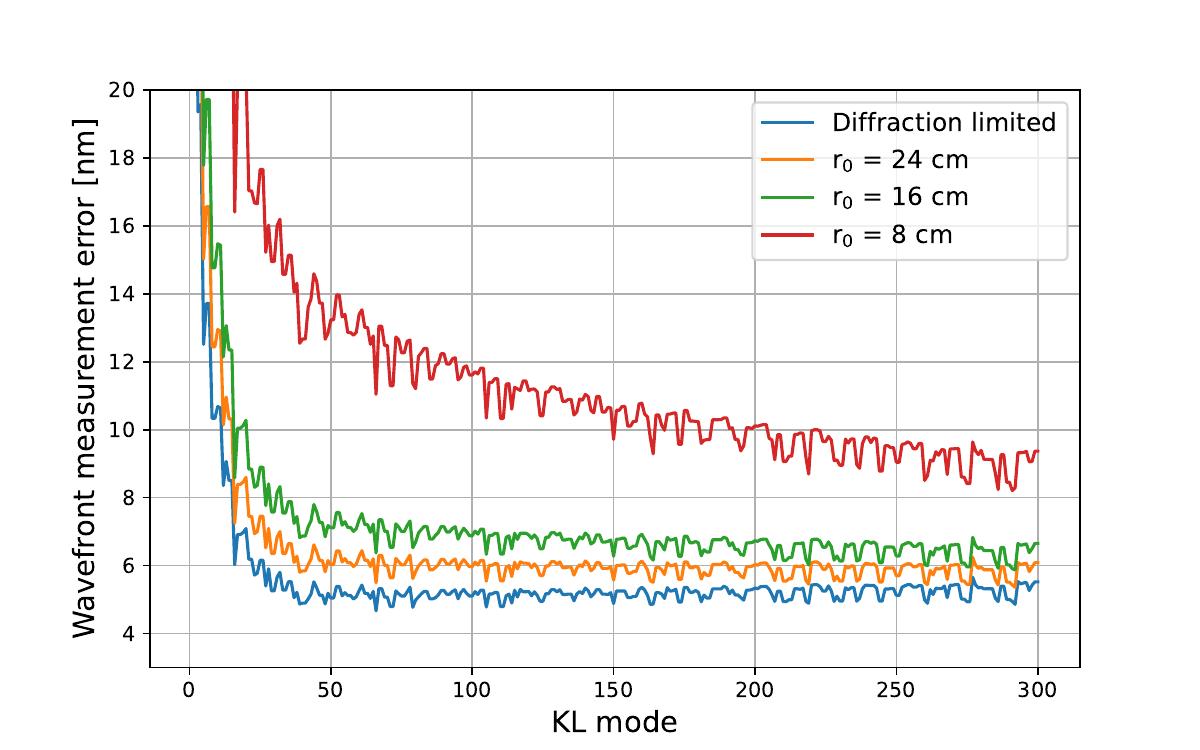}
        \caption{Wavefront measurement error for a Z-band PWFS in different atmospheric conditions (r$_0$).}
        \label{fig:noiseVsKL_diffr0}
    \end{subfigure}
    \caption{PWFS wavefront measurement error due to photon noise for different
    wavelengths and conditions, with $N_{\mathrm{photons}}=1000$, for a modulation of  3\,$\lambda/D$ at Z-band.}
\end{figure}
Using these equations, the expected error on the wavefront measurement due to photon noise is computed, for the diffraction limited case, for different wavelengths, with $N_{\mathrm{photons}}=1000$. The system parameters used in these computations are detailed in Table~\ref{tab:params}. The results are plotted in Fig~\ref{fig:noiseVsKL_diffWaves}, in terms of rms wavefront error. As expected, the noise propagation for the majority of spatial modes is smallest for the shortest wavelength (Z-band). The exception is the lower order modes, where the sensitivity of the PWFS is dominated by the modulation - which in this case has a fixed radius of 3\,$\lambda/D$ at Z-band. To probe the impact of optical gains, the convolutional model proposed by \cite{chambouleyron_pyramid_2020} was used to compute the optical gains due to AO fitting errors for different seeing conditions, which were then applied to the IM. Fig.~\ref{fig:noiseVsKL_diffr0} shows the expected photon noise error for a Z-band PWFS for different conditions. We observe that the diffraction limited case exhibits the smallest noise propagation. As r$_0$ decreases and the AO-corrected phase residuals increase, the PWFS becomes less sensitive and the propagated noise increases.

\begin{figure}
    \centering
        \centering
        \includegraphics[width=0.98\linewidth]{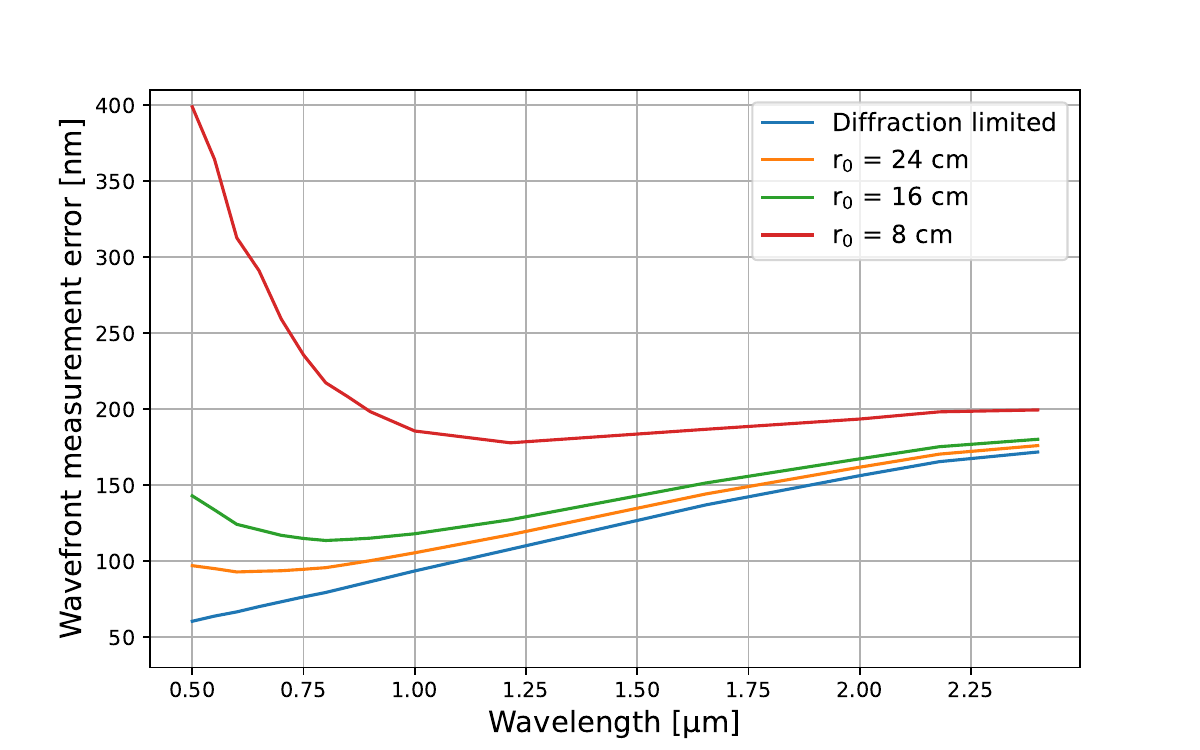} 
    \caption{Total PWFS rms measurement error due to photon noise 
    versus wavelength for different seeing conditions,
    with $N_{\mathrm{photon}}=1000$ and for a modulation radius  of 3\,$\lambda/D$ at $\lambda=$\,500\,nm.}
    \label{fig:noiProp}
\end{figure}

Finally, we compute the total propagated noise by summing the noise variance over the control modes. The results, in terms of rms wavefront error, as a function of wavelength, are shown in  Fig.~\ref{fig:noiProp}. The total propagated noise behaves as expected for the diffraction-limited case - the shortest wavelengths exhibit the least noise (highest signal-to-noise ratio) whilst the noise is greatest for the longest wavelengths. However, as the seeing degrades (smaller $r_0$), the sensitivity decreases and we observe an increase in the propagated noise, with a greater impact at shorter wavelengths where the optical gain effect is greater. For smaller $r_0$ the effect is more pronounced, resulting in a wavelength-dependent noise minima occurring at longer wavelengths as the observing conditions deteriorate. The shorter wavelength sensitivity advantage is negated by this deterioration.

\subsection{Polychromatic reconstruction}

 Using a polychromatic PWFS and the equations outlined in section~\ref{sec:noiseProp}, we can weight the modal measurements at different wavelengths to minimise the noise propagation for each mode. This weighting will depend on the number of photons per band and a measurement of the optical gain, which we will present in section~\ref{sec:OG_tracking}.

The polychromatic reconstruction is computed as:
\begin{equation}
    c_{\mathrm{poly}}(\phi_i) = \sum_{\lambda=1}^{N} w_{\lambda}(\phi_i)c_{\lambda}(\phi_i),
\end{equation}

where $c_{\mathrm{poly}}(\phi_i)$ are the polychromatic modal coefficients, $c_{\lambda}(\phi_i)$ are the modal coefficients at each individual wavelength and $w_{\lambda}(\phi_i)$ are the reconstruction weights for each wavelength. Without additional information, we might consider taking the mean of the reconstruction at different wavelengths, in which case $w_{\lambda}(\phi_i) = 1$. However, using the equations outlined in the previous section, we can use our knowledge of the noise propagation at each wavelength to minimise the polychromatic noise propagation. In this case, the weights are computed as:
\begin{equation}
    w_{\lambda}(\phi_i) = \frac{[\sigma^{2}_{\lambda}(\phi_i)]^{-1}}{\sum \limits_{\lambda=1}^{N} [\sigma^{2}_{\lambda}(\phi_i)]^{-1}},
\end{equation}

where $\sigma^2_{\lambda}(\phi_i)$ are the phase variances for mode $\phi_i$ at wavelength $\lambda$. 

\begin{figure}
    \begin{subfigure}[t]{1\linewidth}
        \centering
        \includegraphics[width=0.98\linewidth]{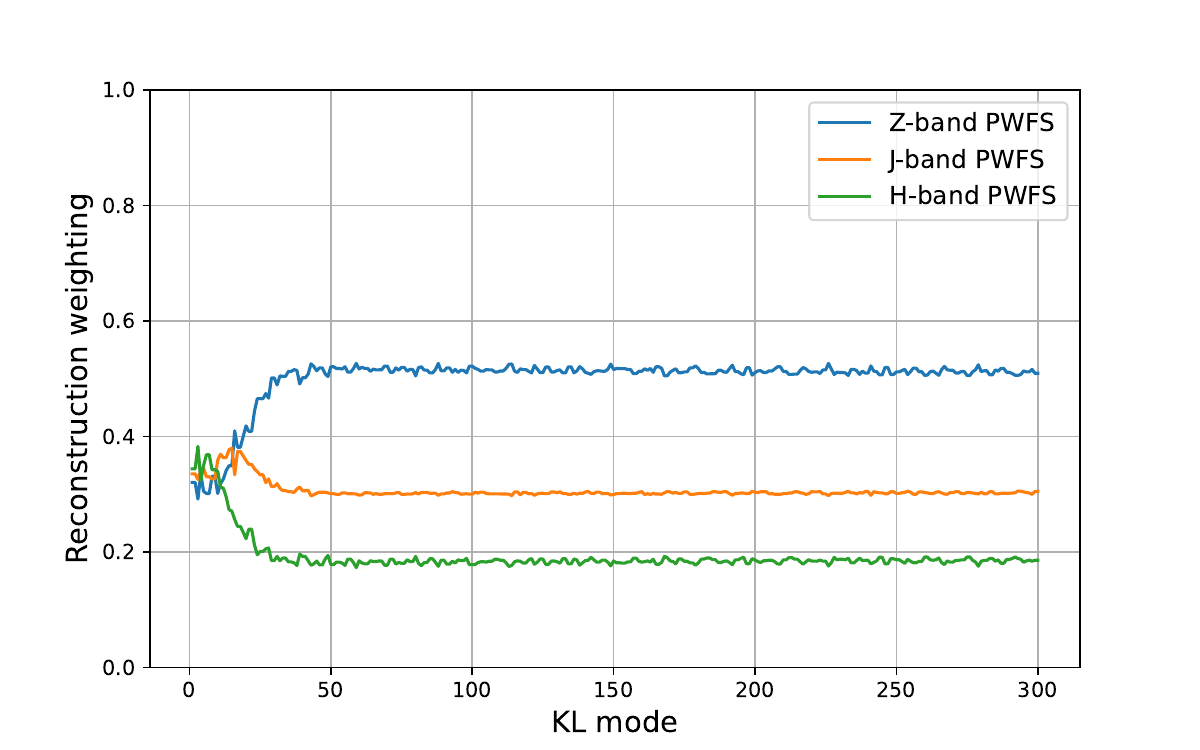}
        \caption{Polychromatic reconstruction weights for each band.}
        \label{fig:polyPWFS_weight}
    \end{subfigure}
    \begin{subfigure}[t]{1\linewidth}
        \centering
        \includegraphics[width=0.98\linewidth]{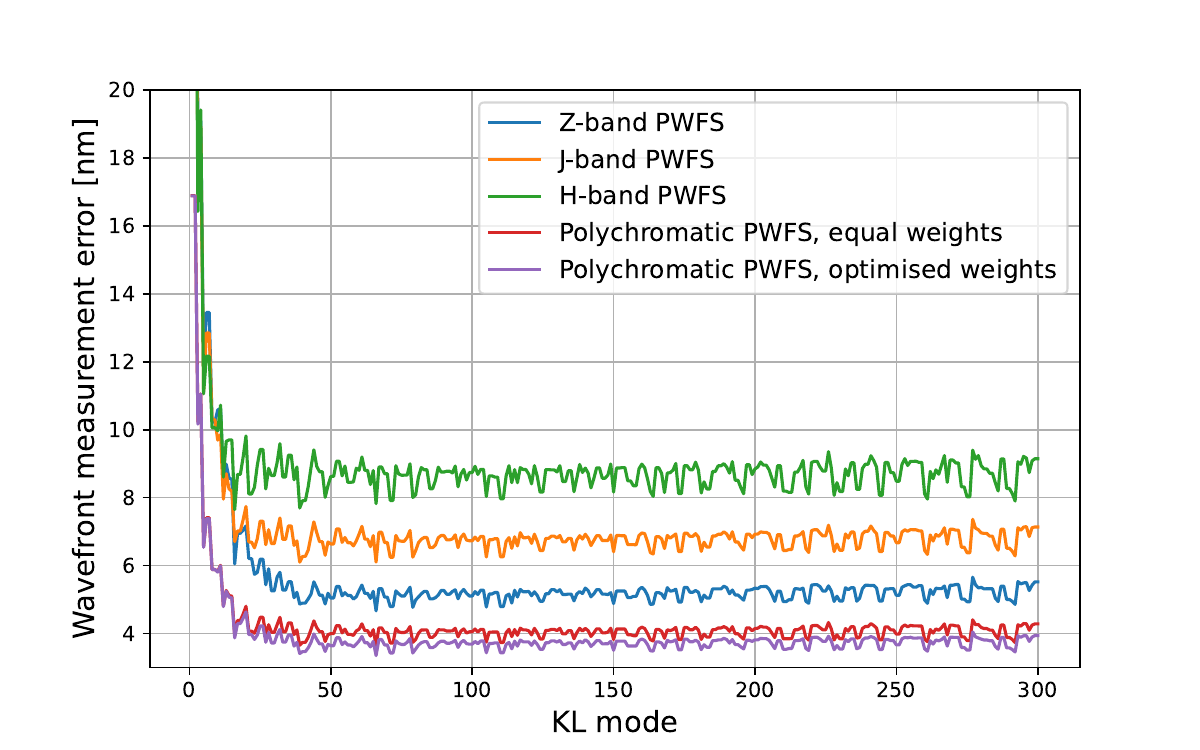}
        \caption{The rms wavefront error due to photon noise for single band and polychromatic PWFSs.}
        \label{fig:polyPWFS_noise}
    \end{subfigure}
    \caption{Polychromatic reconstruction weights and wavefront errors for a diffraction limited polychromatic PWFS, with a modulation of 3\,$\lambda/D$ at Z-band, with $N_{\mathrm{photons}}=1000$ per waveband.}
    \label{fig:polyRecon}
\end{figure}

Fig.~\ref{fig:polyRecon} shows an example of the optimised reconstruction weights (Fig.~\ref{fig:polyPWFS_weight}), and resulting noise propagation (Fig.~\ref{fig:polyPWFS_noise}) for a diffraction limited polychromatic PWFS using three bands (Z-band, J-band and H-band), with $N_{\mathrm{photons}}=1000$ per waveband. For most modes, the optimised reconstruction is weighted towards the shorter wavelengths, as the noise propagation is lower at these wavelengths (see Fig.~\ref{fig:noiseVsKL_diffWaves}). The impact of modulation dominates the sensitivity of the lower order modes and so these weights are roughly equal. Fig.~\ref{fig:polyPWFS_noise} shows the rms wavefront error due to noise for a PWFS operating at each band individually and for a polychromatic PWFS with different reconstruction weights. The combination of the signals at different wavelengths result in a reduction in noise propagation, which is further reduced using the optimised weights (compared to equal weights). The results shown here represent a PWFS working with a diffraction limited PSF with an equal number of photons per waveband. Different distributions of photons across the wavebands and the optical gain effect (operation with a non-diffraction limited PSF) will require modifications to the weights. To optimise the weights during operation requires knowledge of the number of photons in each band as well as an estimate of the optical gain. Optical gain tracking will be the focus of the next section.
  
\section{Polychromatic optical gain tracking method}
\label{sec:OG_tracking}

This section describes the method for performing optical-gain tracking using chromatic wavefront measurements. It first explains how the measurements are mathematically processed, and then outlines the successive steps of the algorithm.
\subsection{Optical gain ratios}

Considering Eq.~\ref{eq:OG}, it is possible to compare the different measurements of the wavefront at each wavelength,  when using a polychromatic PWFS . We define ${\Theta_{og}}( \lambda, \phi_i) $ as the ratio of the optical gain for mode $\phi_i$  as measured at wavelength $\lambda$ compared to the measurement of $\phi_i$ at  reference wavelength $\lambda_0$: 
\begin{equation}
{\Theta_{og}}( \lambda, \phi_i)  = \frac{{g}_{opt}(\lambda, \phi_i)}{{g}_{opt}(\lambda_0, \phi_i)}  =  \frac{c_{\lambda}(\phi_i)}{c_{\lambda_0}(\phi_i)}
\end{equation}
 where $c_{\lambda}(\phi_i)$ and $c_{\lambda_0}(\phi_i)$ are the reconstructed measurements of $\phi_i$ at $\lambda$ and $\lambda_0$. 
In our approach the reference wavelength is defined as the longest wavelength, as it is the least impacted by optical gains. Given this the ratios for shorter wavelengths are expected to be less than one.

Considering the uncertainty on the wavefront measurement, using the instantaneous ratio of the reconstructed modes introduces significant errors (e.g. when the observed mode coefficient approaches zero). Instead, we consider $\Theta_{og}(\lambda, \phi_i)$ as the linear fit between  a series of measurements of  $c_\lambda(\phi_i)$ and $c_{\lambda_0}(\phi_i)$. As all the wavefront measurements at the different wavelengths are used, using a least square fit between $c_{\mathrm{poly}}(\phi_i)$ and both $c_\lambda(\phi_i)$ and $c_{\lambda_0}(\phi_i)$ is more robust. The equation to track the ratio over a time $\delta t$ becomes:
\begin{equation}
     \Theta_{og}(\lambda, \phi_i, \delta t) = \frac{LSQ(c_{\mathrm{poly}}, c_\lambda)}{LSQ(c_{\mathrm{poly}}, c_{\lambda_0})},
\label{eq:least_square_ratio}
\end{equation}
where $LSQ(c_{\mathrm{poly}}, c_\lambda)$ is the least square fit between $c_{\mathrm{poly}}(\phi_i, t) $ and $c_{\lambda}(\phi_i, t) $ defined by
\begin{equation}
LSQ(c_{\mathrm{poly}}, c_\lambda) =
\frac{%
  \begin{aligned}[t]
  &\sum_{t=t_0}^{\delta t} 
  \Big[\Big( c_{\mathrm{poly}}(\phi_i, t) - \langle c_{\mathrm{poly}}(\phi_i, t)\rangle \Big) \\
  &\hspace{60pt}\times \Big( c_\lambda(\phi_i, t) - \langle c_\lambda(\phi_i, t)\rangle \Big)\Big]
  \end{aligned}
}{\begin{aligned}[t]
  \sum_{t=t_0}^{\delta t} 
  \Big( c_{\mathrm{poly}}(\phi_i, t) - \langle c_{\mathrm{poly}}(\phi_i, t)\rangle \Big)^2
  \end{aligned}
},
\label{eq:LSQ}
\end{equation}
and where $LSQ(c_{\mathrm{poly}}, c_{\lambda_0})$ is the least square fit between $c_{\mathrm{poly}}(\phi_i, t) $ and $c_{\lambda_0}(\phi_i, t)$.

To increase the SNR and accelerate the processing the ratios are averaged over the radial order of the modes $n_{th}$:

\begin{equation}
  n_{th}\, \mathrm{order} \Rightarrow  \mathrm{KL \,modes} \in \left[ \frac{n(n+1)}{2} , \,\frac{(n+1)(n+2)}{2} -1 \right],
  \label{eq:radial_order}
\end{equation}
assuming that the indexing starts at 0 (i.e. the first mode corresponds to index 0).

\subsection{On-sky optical gain tracking}

We detail here the methodology used to track optical gains through the wavefront chromatic measurements, outlining each step.

\textit{1. Optical gain look up table generation.}  To determine the theoretical optical gains, we generate a look-up table of computed optical gains for varying $r_0$ and the different central wavelengths of each colour band of the polychromatic PWFS (e.g. Fig. \ref{fig:OG_at_various_wavelengths}). These are generated using the convolutional model described by \citep{chambouleyron_pyramid_2020} and a reconstructed PSF taking into account the fitting error of the system, as described in the previous section. We then compute the optical gain ratios with respect to a reference wavelength (typically the longest wavelength).

\textit{2. Optical gain ratio from chromatic wavefront measurements.}  During on-sky operation, we use
Eq.~\ref{eq:least_square_ratio} to provide the optical gain ratios measured from a set of polychromatic PWFS signals with respect to the reference wavelength. 

\begin{figure}
    \centering
    \includegraphics[width=.98\linewidth]{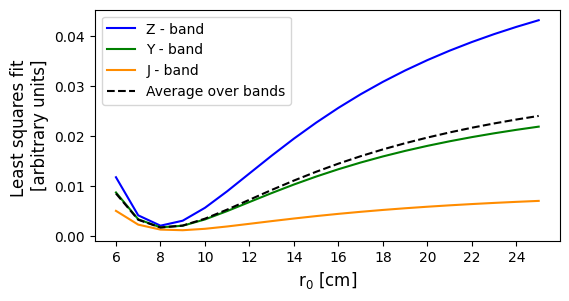}
    \caption{Closed loop data ratios fit to the optical gain look-up table ratios, parametrized by $r_0$. The ratios are Z, Y and J bands data over H band data, for a simulated closed loop with $r_0=8$\,cm. The measured value of $r_0$ is given by the minimal value of the average of the fits.}
    \label{fig:fit_ratio}
\end{figure}

\textit{3. Optical gain ratio fitting.}
We compute the least squares fit  between the measured $\Theta_{og}$ over time and the optical gain ratios computed from the look-up table. This difference is averaged over all the PWFS colour bands. The best fit between the model and the measured $\Theta_{og}$ returns an estimate for $r_0$,  as shown in Fig. \ref{fig:fit_ratio}, and the associated optical gains from the table. 

\textit{4. Optical gain compensation.}
The optical gain is compensated in the closed loop by applying a gain correction factor of $1/g_{opt}(\lambda,\phi_i)$ to the reconstruction.


For a high flux case operating at high speeds, where we expect the optical gain to be dominated by the fitting error, the measured value of $r_0$ is close to the true value. However, with additional wavefront errors present in the residual, such as noise and temporal errors, we expect the value of $r_0$ to be underestimated as the theoretical optical gains for smaller $r_0$  better fit the measured data. As the goal here is to track the optical gain this is not critical. Future work could adapt the look up table to include noise and AO frame rate as additional parameters. The impact of noise is studied in section~\ref{sec:OG_tracking_noise}.

\section{Simulation study}
\label{sec:simulation_study}

\subsection{Reference Design: Keck II Adaptive Optics System}
\label{sec:ao_system_keck}

We have selected a reference system to aid in evaluating the polychromatic PWFS performance improvements and practical implementation issues to be discussed in subsequent sections.

\begin{figure}
    \centering
    \includegraphics[width=1\linewidth]{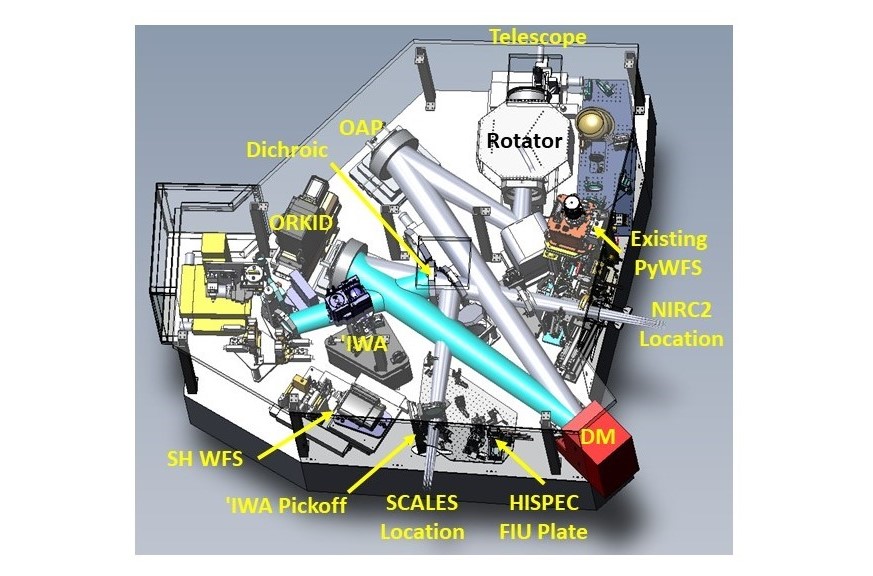}
    \caption{Keck II AO bench. Light from the telescope passes through a K-mirror rotator to a fast tip-tilt mirror and off-axis parabola (OAP) to the deformable mirror (DM) and a second OAP. An IR transmissive dichroic then reflects the visible light and transmits the longer wavelength light. A fold mirror reflects the light toward SCALES and HISPEC. The 'IWA pickoff reflects light to the 'IWA location while transmitting light to SCALES and HISPEC.}
    \label{fig:AOB}
\end{figure}

Our reference system is the Keck II AO system, with a 2844-actuator deformable mirror, corresponding to 20~cm subapertures on the Keck telescope primary mirror. Our performance reference is the planned near-infrared PWFS system identified as 'IWA (Infrared Wavefront sensor for Astronomy) \citep{wizinowich_adaptive_2024, lilley_keck_2024} in Fig. \ref{fig:AOB}. The polychromatic PWFS would replace 'IWA at this location. The science instruments to be supported by 'IWA or the polychromatic PWFS are HISPEC \citep{mawet_fiber-fed_2024}, ORKID \citep{peretz_orcas_2025} and SCALES \citep{stelter_scales_2024}. The current IR transmissive dichroic, which transmits  $\lambda >$ 950 nm, would be replaced by one that transmits $\lambda >$ 800 nm. This dichroic change could offer potential science range increases for HISPEC and SCALES at the expense of ORKID, as well as the visible WFSs that use light reflected by this dichroic.

This study is performed for a polychromatic PWFS operating from 800 nm to 1800 nm. This range was selected since (1) MKIDs are already capable of handling this wavelength range, (2) longer wavelengths introduce thermal emission design issues and are also key science wavelengths, and (3) shorter wavelengths experience significant atmospheric dispersion, AO correction is poor, and these wavelengths are used by other sensors.

The polychromatic PWFS is expected to operate in the following wavelength bands  (as summarized in Table \ref{tab:PWFS_bands}):
\begin{enumerate}
    \item 800 nm to 1800 nm for HISPEC and/or SCALES science at K- to M-band, or ORKID science less than 800 nm. This requires a dichroic (the 'IWA pickoff in Fig. \ref{fig:AOB}) reflecting 800 nm to 1800 nm light to the PWFS and transmitting $\ > $  1800 nm light to HISPEC and SCALES.
    \item 800 nm to 1100 nm for HISPEC and/or SCALES science at J- to M-band. This requires a dichroic reflecting 800 nm to 1100 nm light to the PWFS and transmitting$\ > $ 1100 nm light to HISPEC and SCALES.
    \item 1100 nm to 1800 nm for HISPEC and/or SCALES science at Z and Y-band; requiring a dichroic reflecting 1100 nm to 1800 nm light to the PWFS and transmitting < 1100 nm light to HISPEC and SCALES.
    \item Other intermediate options could also be considered. 
\end{enumerate}

\begin{table}
    \centering
    \caption{Wavelengths (nm) sent to the polychromatic PWFS and science intruments by choice of dichroic beamsplitter. }
    \begin{tabular}{c c c c }
    \hline
      \textbf{Option} & \textbf{PWFS} & \textbf{HISPEC/SCALES} & \textbf{ORKID}\\
      
    \hline
    (i) & 800-1800 & > 1800 & < 800 \\
    (ii) & 800-1100 & > 1100 & < 800 \\
    (iii) & 1100-1800 & < 1100 & < 800 \\
    
    \hline
    \end{tabular}

    \label{tab:PWFS_bands}
\end{table}

\begin{figure}
    \centering
    \includegraphics[width=.9\linewidth]{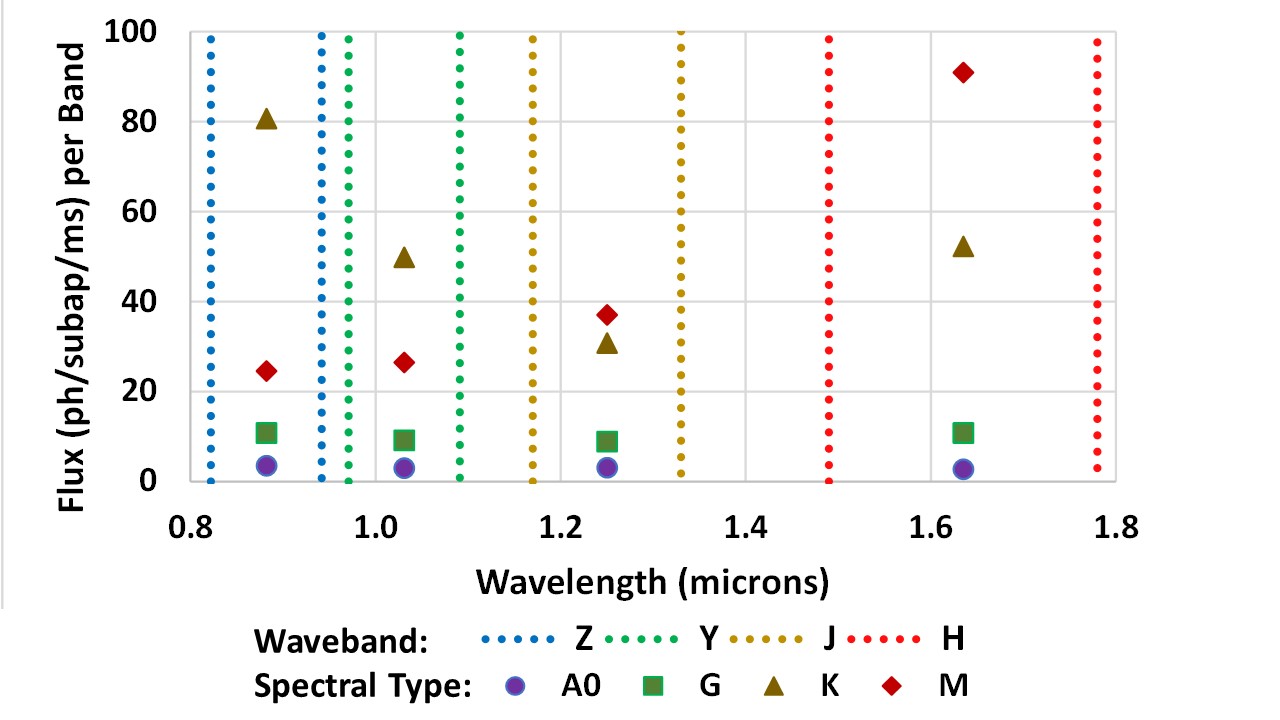}
    \caption{Flux (photons/subaperture/ms) incident on the detector in each waveband (Z, Y, J and H) versus spectral type (A0, G, K and M); assuming a v = 10 magnitude star, 20~cm square subapertures and 0.25 system throughput (based on \protect\cite{dekany_wavefront_2024}). The total flux available for wavefront sensing increases by factors of 2.0 to 4.5, depending on spectral type, by using all four wavebands instead of just H-band. Note that atmospheric absorption is significant between these wavebands.}
    \label{fig:Flux}
\end{figure}

Fig. \ref{fig:Flux}. is a plot of the flux at the detector from a star of a given spectral type (A0, G, K and M) in the Z, Y, J and H-bands, within the 800 nm to 1800 nm band.

\begin{figure}
    \centering
    \includegraphics[width=0.8\linewidth]{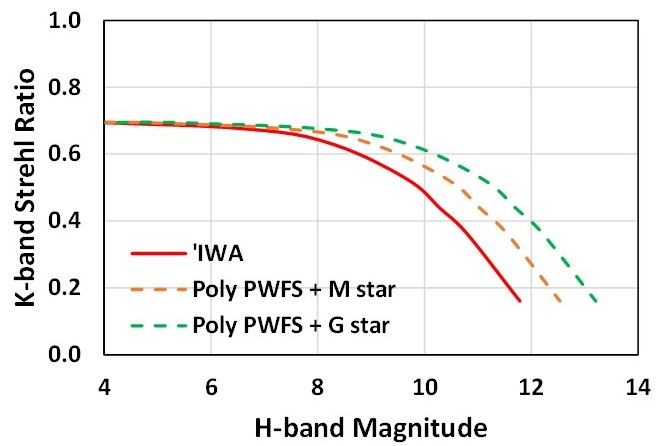}
    \caption{Keck II AO predicted Strehl ratio when using 'IWA in H-band versus a similar polychromatic PWFS using Z, Y, J and H-band for two different spectral type (G and M) stars. The polychromatic PWFS improvement in limiting magnitude is just due to the additional photons from the wider passband and assumes perfect optical gain correction for the shorter wavelength.}
    \label{fig:IWA Strehl}
\end{figure}

Fig. \ref{fig:IWA Strehl}. shows 'IWA's predicted K-band Strehl ratio performance with 'IWA operating in the H-band versus a polychromatic PWFS operating using Z to H-band; based on the measured on-sky Keck II AO performance with a lower order H-band PWFS \citep{bond_adaptive_2020}. 'IWA can operate with H- or J-band light reflected by a choice of two narrow-band reflective dichroics. A polychromatic PWFS could extend performance to 0.75 to 1.6 magnitude fainter stars, depending on spectral type, just due to the increased photon flux from using Z to H-band. The associated sky coverage increase will be on the order of factors of 2 to 4 \citep{robin_synthetic_2003}. This result does not take into account the differences in noise propagation at different wavelengths or optical gain effects at shorter wavelengths.

For the bright star case, the Strehl ratio on the polychromatic PWFS would vary from 0.1 at 800 nm to 0.6 at 1800 nm. This wavelength-dependent image quality variation would need to be accounted for in how the the polychromatic WFS is calibrated, how gains are calculated on sky, and how different wavelength information might be weighted for AO correction.

\subsection{Simulation framework}

To understand the system and explore the potential benefits of polychromatic wavefront sensing, it is important to model an example of such a system. The requirements of the MKID devices needed for different AO systems can also be defined.

An end-to-end simulation of a single-conjugate AO system using a PWFS for a 10 m telescope was carried out with the Object-Oriented Python AO package (OOPAO; \cite{heritier_oopao_2023}). The simulation configuration is outlined in Table~\ref{tab:params}. To replicate a polychromatic sensor, individual natural guide star and PWFS components are created for each designated colour band. Here, each band is modelled using a single representative wavelength. We initially consider performance in the absence of photon noise.

To recreate the effect of a single modulation mirror, the modulation radius in terms of $\lambda/D$ is scaled by the wavelength. Wavefront sensing can be performed either separately for each wavelength (single-band scenario) or by integrating simultaneous measurements across multiple wavelengths (polychromatic scenario). For multi-wavelength operation, the modulation in units of $\lambda/D$ is defined relative to the first wavelength listed, which is generally the shortest.

\begin{table}
    \centering
    \caption{Simulation parameters. The modulation radius is chosen for $\lambda$ being the shortest wavelength of the series of wavelengths for each simulated case.}
    \begin{tabularx}{\linewidth}{ >{\arraybackslash}X 
                               >{\arraybackslash}X }
    \hline
        \textbf{ Parameter    }      & \textbf{Value}             \\
        
         \hline
        
         Telescope diameter     & 10\,m              \\
         Simulation resolution  & $200\times 200$ pixels\\
         PWFS sampling           & $50\times 50$ pixels\\
         Modulation radius& $3~\lambda/D$\\
         Number of controlled modes& 500                   \\
         Frame rate             & 500\,Hz                \\
         Delay                  & 2 frames              \\
         Average wind speed     &  11\,m/s                    \\
         \hline
    \end{tabularx}

    \label{tab:params}
\end{table}

\subsection{Colour bands distribution}
\label{sec:colour_bands}

To choose the different wavebands at which the wavefront will be reconstructed, there are three points to consider: the bandpass left available by the science instrument, the energy resolution achievable by the detector, and the number of bands required. Currently, MKID arrays demonstrate energy resolution at 400 nm between 20 \citep{walter_mkid_2020} for large arrays and 50 \citep{de_visser_phonon-trapping-enhanced_2021} for individual pixel devices. 

The first colour configuration considered is determined using the energy resolution already demonstrated on-sky by the MEC instrument, which uses a large MKID array of the kind that would be required for a PWFS. To compute the energy resolution of a detector, $R(\lambda$), at a given wavelength $\lambda$ we use the following equation:
\begin{equation}
    R (\lambda) = R_0 \times \frac{\lambda}{\lambda_0},
    \label{eq:energy_res}
\end{equation}

where $R_0$ is the energy resolution at the reference wavelength $\lambda_0$. We then compute the colour bands achievable within the 800 to 1800 nm bandwidth. The results are presented Table  \ref{tab:Case_1}.

Considering the waveband from 800~nm to 1800~nm and referring to Fig.~\ref{fig:Flux}, the following colour band configurations are defined for the simulation case study aimed at applying the polychromatic optical gain tracking method:

\begin{itemize}
    \item Case 1: All the colour bands presented in Table~\ref{tab:Case_1}
    \item Case 2: Z, Y, J, and H bands
    \item Case 3: two colour groups: Z and Y bands; J and H bands
    \item Case 4: Z and Y bands
    \item Case 5: J and H bands
\end{itemize}

Case~1 is designed to maximise the number of bands that can be resolved with an MKID array having an energy resolution similar to the MEC array. Case~2 corresponds to the band structure shown in Fig.~\ref{fig:Flux}, optimised to account for atmospheric absorption features. Case~3 groups the bands from Case~2 into pairs, allowing an assessment of the performance impact when reducing the number of colour bands across the same spectral range—a strategy that could be employed to compensate for low signal levels. Finally, Cases~4 and~5 are tailored for K- and M-type stars, whose flux is predominantly concentrated in the Z/Y and J/H bands, respectively.

Table \ref{tab:rms_og_colours} presents the rms error between the optical gains computed at the central wavelength of each colour band and those computed at the edges of the Z, Y, J, and H bands. This error represents the maximum possible deviation in the optical gain measurements due to uncertainties in the photon energy measurements. The rms error is consistently larger at the lower edge of each band, as optical gain variations tend to be more pronounced at shorter wavelengths.

\begin{table}
\centering
\caption{Central wavelength and bandwidth data for the first colour configuration case. A spectral resolution of $R_0$ = 19 was used to sufficiently define these bands.}
\begin{tabularx}{\linewidth}{>{\centering\arraybackslash}X 
                             >{\centering\arraybackslash}X 
                            >{\centering\arraybackslash}X 
                               >{\centering\arraybackslash}X }
                              
\hline
 \textbf{\# Colour} & \textbf{Central $\lambda$ } & \multicolumn{2}{c}{\textbf{Colour edges (nm)}} \\ 
 \cline{3-4}
 \textbf{band} & \textbf{(nm)}& \textbf{Start} & \textbf{End} \\
\hline

 1 & 842& 800     & 887\\
 2 & 933& 887& 988\\
3 & 1044& 988& 1113\\
 4 & 1183& 1113& 1272\\
 5 & 1361& 1272& 1479\\
 6 & 1598& 1479& 1760\\
\hline

\end{tabularx}

\label{tab:Case_1}
\end{table}

\begin{table}
\centering
\caption{The rms error in optical gain estimation when measured at the minimum or maximum wavelength of each color band, for Fried parameter values of $r_0 = 8$~cm and $r_0 = 16$~cm.}
\begin{tabularx}{\linewidth}{ >{\centering\arraybackslash}X 
                               >{\centering\arraybackslash}X 
                               >{\centering\arraybackslash}X 
                               >{\centering\arraybackslash}X 
                               >{\centering\arraybackslash}X }
                        \hline      
                \textbf{Colour}        & \multicolumn{2}{c}{\textbf{$r_0 = 8$~cm}}   & \multicolumn{2}{c}{\textbf{$r_0 = 16$~cm}}  \\ \cline{2-5} 
                 \textbf{Band}       & \textbf{Min $\lambda$} & \textbf{Max $\lambda$} & \textbf{Min $\lambda$} &\textbf{ Max $\lambda$}  \\ \hline
\multicolumn{1}{c}{Z} & 0.1261         & 0.0548          & 0.0579         & 0.0239          \\
\multicolumn{1}{c}{Y} & 0.0954         & 0.0418          & 0.0402         & 0.017           \\
\multicolumn{1}{c}{J} & 0.0873         & 0.0372          & 0.0341         & 0.0141          \\
\multicolumn{1}{c}{H} & 0.0749         & 0.0287          & 0.0267         & 0.01            \\ \hline
\end{tabularx}
\label{tab:rms_og_colours}
\end{table}

\subsection{Optical gain tracking performance}
\label{sec:OG_tracking_perf}

To evaluate the performance of polychromatic optical gain tracking, we performed a series of end-to-end simulations of AO closed-loop scenarios using the parameters shown in Table {\ref{tab:params}}. Five different colour band configurations (Cases 1–5, as defined in section \ref{sec:colour_bands}) were tested under two seeing conditions corresponding to Fried parameters $r_0=8$~cm and 16~cm. The performance metric is the final K-band Strehl ratio achieved after AO correction.

\begin{figure}
    \centering
    \includegraphics[width=.98\linewidth]{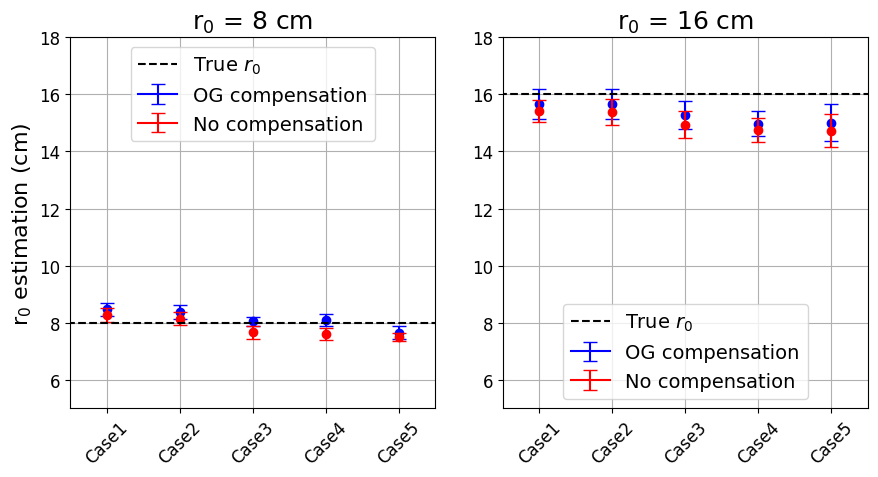}
    \caption{Comparison of the atmospheric conditions tracking at $r_0 = 8$~cm and $r_0 = 16$~cm using different colour configuration cases. Error bars are computed using the standard deviation of the measurements at 500 Hz over 4 seconds of data.  }
    \label{fig:r0_vs_cases}
\end{figure}

Fig.~\ref{fig:r0_vs_cases} presents the average tracking of the $r_0$ value over a 4 second period, using data acquired at a 500~Hz frame rate for the target $r_0$ values. At $r_0$ = 8~cm, the optical gain is significant, resulting in a highly stable $r_0$ estimate. At 16~cm, discrepancies begin to emerge between the performances of different colour configuration cases. Specifically, configuration cases~1 and~2, which utilize six and four spectral bands respectively, demonstrate superior performance compared to configurations~3, 4 and~5, which rely on only two bands. This shows the advantage of employing more than two spectral measurements to stabilise accurate tracking when the method is less sensitive.

\begin{figure}
    \centering
    \includegraphics[width=.98\linewidth]{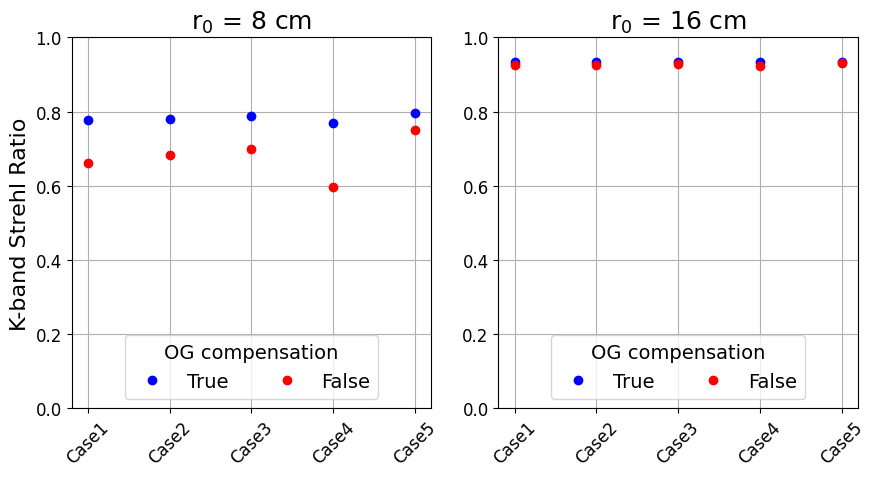}
    \caption{Comparison of Strehl ratio for different colour band cases described in section~\ref{sec:colour_bands}, under atmospheric conditions $r_0 =$ 8~cm and $r_0=$ 16~cm.}
    \label{fig:strehl_vs_cases}
\end{figure}

Fig. \ref{fig:strehl_vs_cases} compares the K-band Strehl ratios for each case with and without polychromatic optical gain tracking. Across all cases, the use of optical gain tracking significantly improved performance in poor seeing conditions. Optical gain effects have a smaller impact when residual phase aberrations are small (e.g. at longer wavelength as evidenced in Fig. \ref{fig:Chromatic_OG_variation}) and no significant improvement was obtained in these simulations at $r_0 = 16$~cm, although we note that these are photon noise-free simulations. 

Both cases 4 and 5 use only part of the full 800 to 1800~nm available bandwidth, with Case 4 using only the shorter colour bands, and case 5 the longer. Comparing these two cases in Fig. \ref{fig:strehl_vs_cases} demonstrates the improvement in correcting for the optical gain, using the proposed polychromatic PWFS, can bring significant benefits to PWFS operating at shorter wavelengths. For PWFS operating at longer wavelengths, optical gain compensation does not introduce such significant performance improvements in terms of Strehl ratio.

\begin{figure}
    \centering
    \includegraphics[width=.98\linewidth]{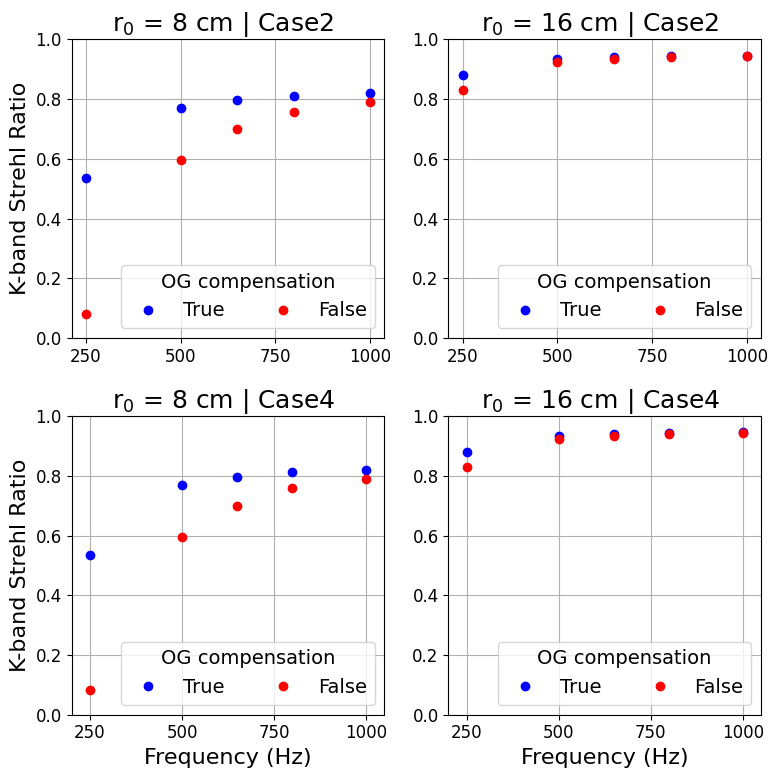}
    \caption{Comparison of Strehl ratio performance for different frame rates of the polychromatic PWFS. It is simulated for the colour configuration cases 2 and 4 and for atmospheric conditions of $r_0 =$ 8~cm and $r_0=$ 16~cm.}
    \label{fig:sr_vs_freq}
\end{figure}

Fig.~\ref{fig:sr_vs_freq} illustrates the impact of varying the PWFS frame rate for Cases~2 and~4. In the absence of optical gain compensation, reducing the frame rate leads to a degradation of the Strehl ratio. This trend is evident in both cases and under both atmospheric conditions considered. When optical gain compensation is applied, the decline in performance is mitigated, particularly for the case of $r_0 = 8$~cm. At a frame rate of 250~Hz, the system performs poorly without compensation, yielding a Strehl ratio below 0.2 for both cases 2 and 4. With compensation, the Strehl ratio remains above 0.5. At $r_0 = 16$~cm, the impact is less pronounced. These results demonstrate that optical gain compensation becomes more critical at the lower loop rates needed to increase the signal-to-noise ratio for fainter guide stars.

\subsection{Optical gain tracking with photon noise}
\label{sec:OG_tracking_noise}

\begin{figure}
    \begin{subfigure}[t]{\linewidth}
        \centering
        \includegraphics[width=0.98\linewidth]{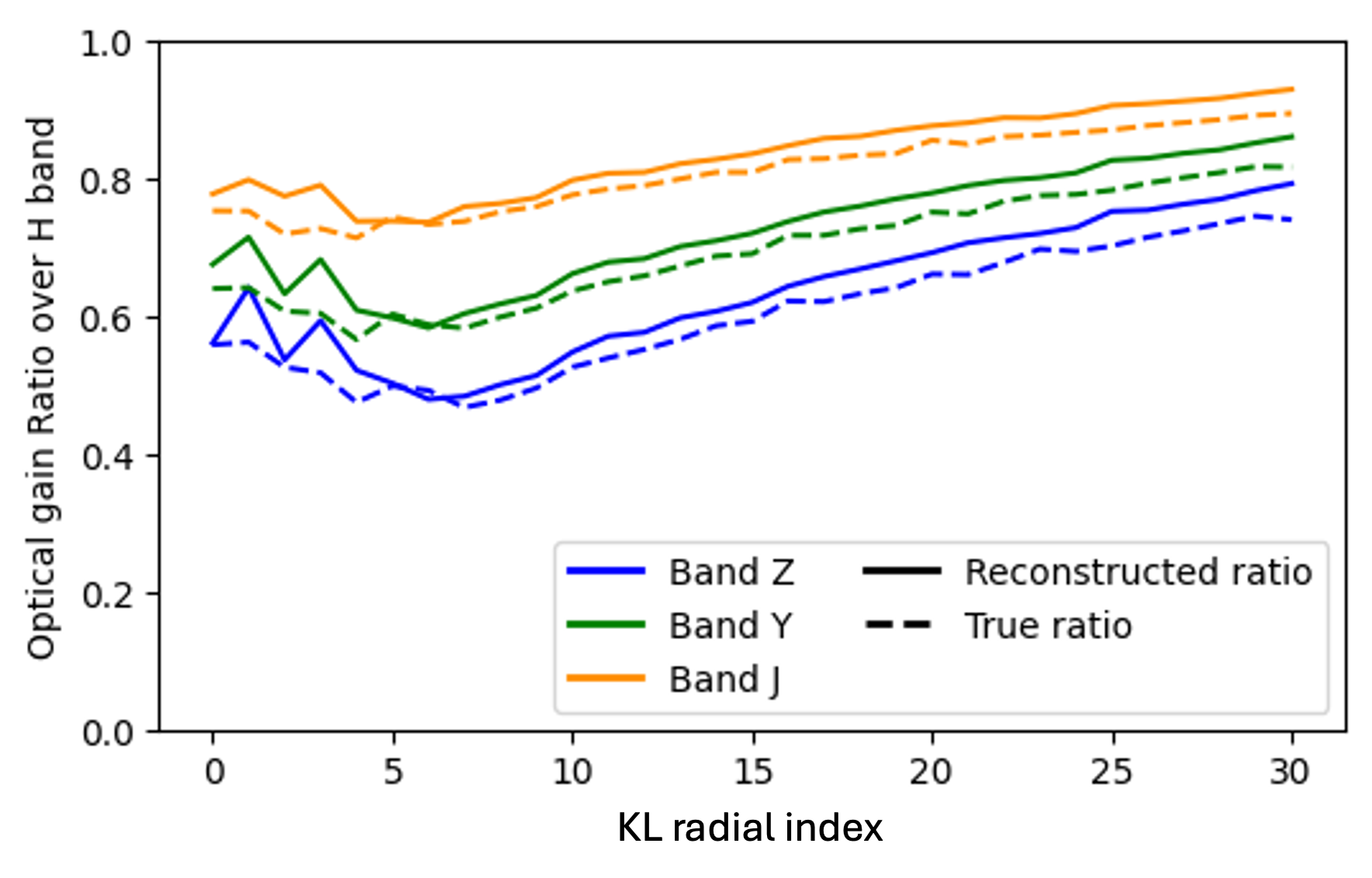}
        \caption{No photon noise}
        \label{fig:LSQNoNoise}
    \end{subfigure}
        \begin{subfigure}[t]{1\linewidth}
        \centering
        \includegraphics[width=0.98\linewidth]{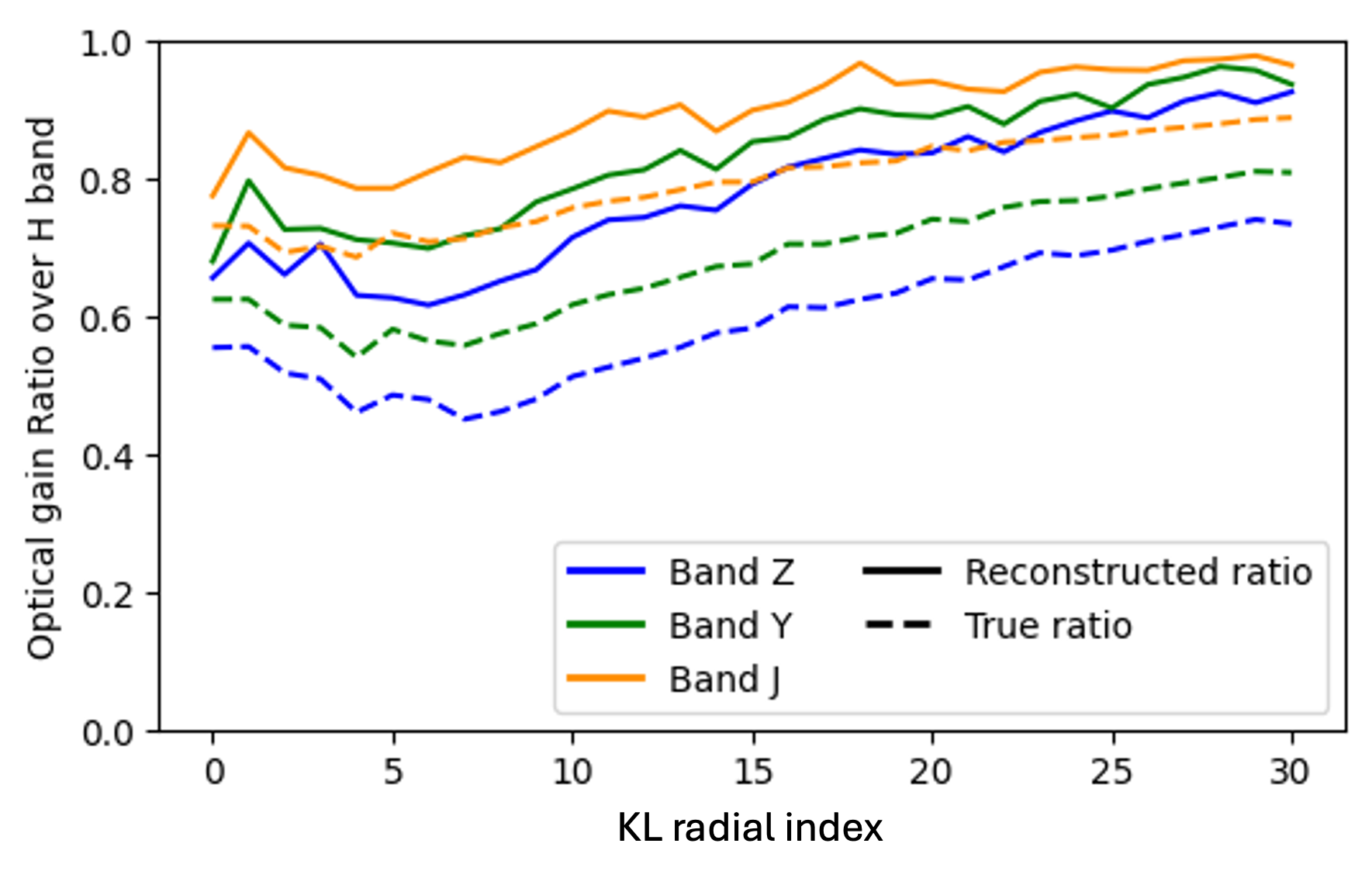}
        \caption{Photon noise for a H = 12.5 magnitude, G-type star}
        \label{fig:LSQNoisy}
    \end{subfigure}   
    \caption{Optical gain ratios reconstructed from 1 second data segments using an end-to-end simulation at a $r_0$ of 8 cm. The true ratios are obtained using the convolutional model for reference. Reconstruction ratios are based on the least-squares method (Eq.~\ref{eq:least_square_ratio}). The integration of the noise in the simulation (b) illustrates the need for a refined method to do the ratios.}
    \label{fig:ratios_methods}
\end{figure}

In the absence of photon noise, the least-squares method (cf Eq.~\ref{eq:least_square_ratio}) is sufficient to estimate the optical gain ratios, as illustrated in Fig.~\ref{fig:LSQNoNoise}. The figure compares the reconstructed ratios with the optical gain ratios obtained from the PSF of the same end-to-end simulation, showing a match between the two. However, the least square approach assumes no noise on the reference wavelength measurement. If there is noise on both the numerator and denominator then the noise will not average out with more data and the optical gain ratios become biased, as shown in Fig.~\ref{fig:LSQNoisy}. The photon level here was chosen to correspond to a G-type star where the impact of photon noise on 'IWA performance is important (i.e. an H-magnitude of 12.5 and Strehl ratio of $\sim$
0.3  in Fig.~\ref{fig:IWA Strehl}). To account for this effect, the noise level must be weighted according to the number of photons detected at each wavelength. For this purpose, we define
\begin{equation}
    \epsilon_\lambda = \frac{\sigma^2_\lambda}{\sigma^2_{\lambda_0}},
\end{equation}
where $\sigma^2_\lambda$ is the noise variance at wavelength $\lambda$, and $\sigma^2_{\lambda_0}$ is the noise variance at a chosen reference wavelength $\lambda_0$. The coefficient $\epsilon_\lambda$ is then used as a weighting factor in the total least-squares formulation, thereby compensating for wavelength-dependent noise \citep{groen_introduction_1998}.
\begin{equation}
    \Theta_{og}(\lambda, \phi_i, \delta t) = \frac{2 \, C_{\lambda,\lambda_0}}{C_{\lambda_0} - \epsilon_\lambda \, C_{\lambda} + \sqrt{(C_{\lambda_0} - \epsilon_\lambda\,C_{\lambda})^2 + 4\,\epsilon_\lambda\,C_{\lambda,\lambda_0}}}
\label{eq:total_least_squares}
\end{equation}

where $C_{\lambda  }$ is the auto-correlation of $c_{\lambda}(\phi_i, t)$ defined by:
\begin{equation}
    C_{\lambda  } = \sum\limits_{t = t_0}^{\delta t} \left( c_{\lambda}(\phi_i, t) - \left< c_{\lambda}(\phi_i, t) \right> \right)^2, 
\end{equation}

\begin{figure} 
    \begin{subfigure}[t]{1\linewidth}
        \centering
        \includegraphics[width=0.98\linewidth]{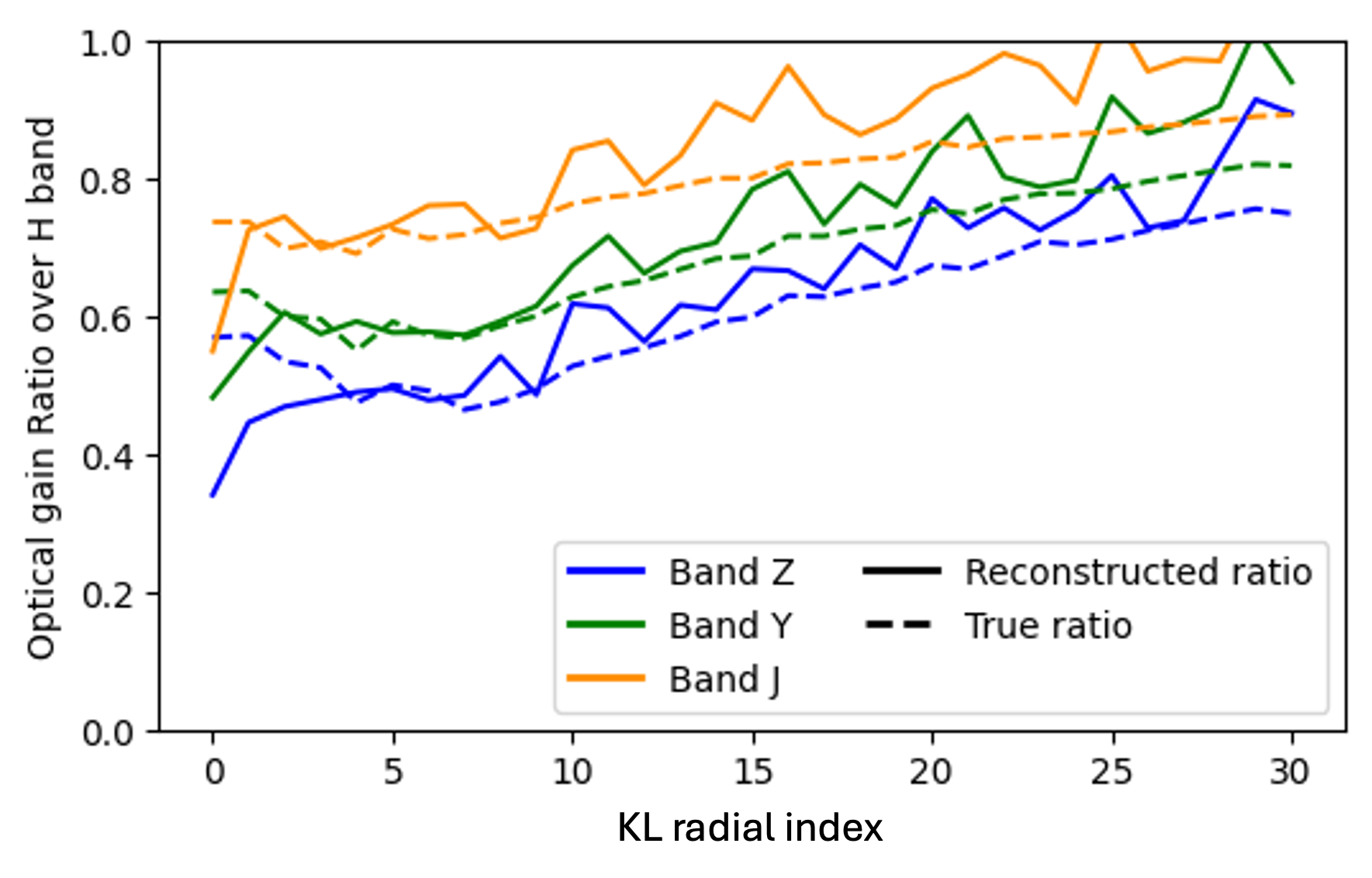}
        \caption{1 second data average}
        \label{fig:TLSQ}
    \end{subfigure}
    \begin{subfigure}[t]{1\linewidth}
        \centering
        \includegraphics[width=0.98\linewidth]{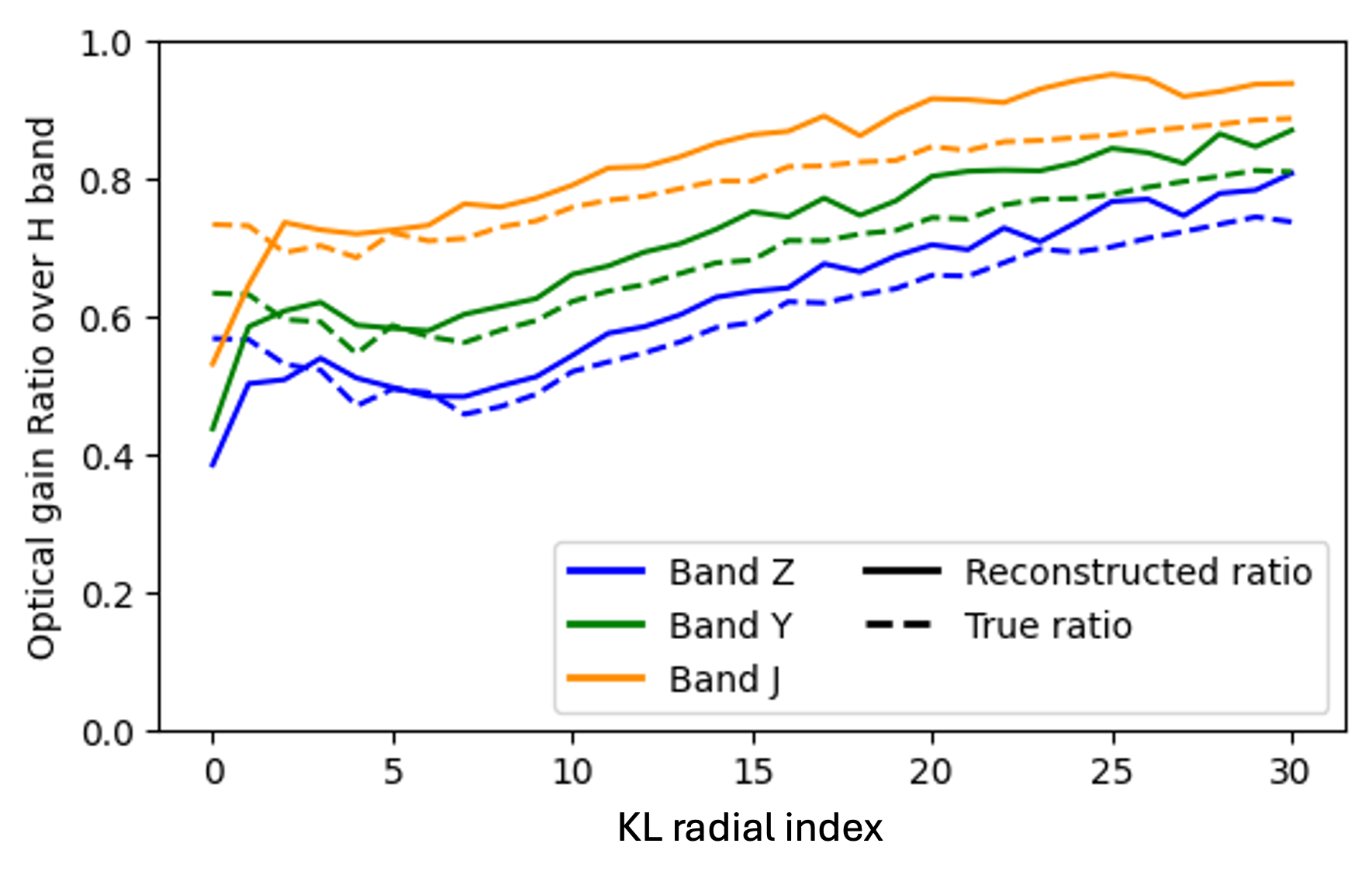}
        \caption{30 second data average}
    \label{fig:TLSQ_30sec}
    \end{subfigure}
    \caption{Optical gain ratios reconstructed from 1 second data \textbf{(a)} and 30 second data \textbf{(b)} using an end-to-end simulation for a H = 12.5 magnitude, G-type star and a $r_0$ of 8 cm. The true ratios are obtained using the convolutional model for reference. Reconstruction ratios are based on the total least-squares method (Eq.~\ref{eq:total_least_squares}) which incorporates noise modelling.}
    \label{fig:ratios_TLSQ_methods}
\end{figure}

$C_{\lambda_0 }$ is the auto-correlation of $c_{\lambda_0}(\phi_i, t)$ and $C_{\lambda, \lambda_0}$ is the correlation between  $c_{\lambda}(\phi_i, t)$  and  $c_{\lambda_0}(\phi_i, t)$ defined by
\begin{multline}
    C_{\lambda, \lambda_0} = \sum\limits_{t = t_0}^{\delta t} \left( c_\lambda(\phi_i, t) - \left< c_\lambda(\phi_i, t) \right>\right) \\ \times \left(c_{\lambda_0} (\phi_i, t) - \left<c_{\lambda_0} (\phi_i, t) \right>\right).
\end{multline}
Fig.~\ref{fig:ratios_TLSQ_methods} shows the ratios using  Eq.~\ref{eq:total_least_squares} over 1 second of data (Fig.~\ref{fig:TLSQ}) and 30 seconds of data (Fig.~\ref{fig:TLSQ_30sec}) and compares them to the optical gain obtained directly from the PSF of the same end-to-end simulation. A comparison with Fig.~\ref{fig:LSQNoisy} highlights the need to account for the noise through the total least squares approach, compared to the standard least squares. The total least squares approach leads to a closer fit with the true optical gain ratios. For the same time interval (1 second) the curves are less smooth using the total least squares due to the total least squares effectively correcting for the noise to measure the underlying variation of the reconstruction with wavelength.

The impact of photon noise can be further mitigated by averaging the signal. While averaging over radial orders is already applied (cf. Eq. \ref{eq:radial_order}), temporal averaging can also be performed, with the integration time adapted to the photon noise level. As shown in Fig. \ref{fig:ratios_TLSQ_methods}, the 30 second dataset is significantly smoother and more closely resembles Fig.~\ref{fig:LSQNoNoise}. This highlights the trade to be considered when implementing the method on-sky: balancing noise mitigation against the update rate of optical gain tracking, which in turn depends on the time scale of $r_0$ variations. The typical time scales for significant $r_0$ variations are greater than 1 minute based on Keck experience \citep{schock_atmospheric_2003, jolissaint_determination_2018}

\subsection{Polychromatic performance improvements}
\label{sec:perf-improve}
In this section, we use end-to-end simulations to assess the performance of a polychromatic PWFS compared to a single band PWFS, over a range of guide star magnitudes. We focus on the performance for a G-star and use a polychromatic PWFS corresponding to Case 2 (4 colours, Z, Y, J and H bands), comparing the performance with Z and H band PWFS (representing the shortest and longest wavelengths). 

When considering how to optimise the performance of an AO system, there are a number of parameters we can tune, such as the WFS detector frame rate, the loop gain and, in the case of the PWFS, the modulation. In this section, the PWFS modulation was fixed at 3\,$\lambda/D$ (at Z band) and a frame rate of 500\,Hz was used. For each case, the modal loop gain was optimised~\cite{gendron_astronomical_1994} to balance the temporal errors against propagated noise, using the theoretical noise computation outlined in section~\ref{sec:noiseProp}. As we only consider photon noise here, we felt it unnecessary to tune the frame rate as well as the loop gain. 

For each case, the optical gain was computed from the closed loop PSF and averaged over the last 200 frames of the loop. For the polychromatic PWFS, this can be achieved using the optical gain tracking method outlined in the previous sections, whereas the single band methods would require alternative methods, either probing modes on-sky or using a focal plane camera to record the PSF incident on the PWFS. When using optical gain compensation, the following control updates are made when the optical gain estimate is updated:
\begin{enumerate}
    \item Optical gain compensation computed as $\frac{1}{g_{opt}}$ for each wavelength
    \item Noise propagation recomputed using the current optical gain estimate (from step i)
    \item Reconstruction weights recomputed for the polychromatic PWFS using the new noise propagation estimate (from step ii)
    \item Modal loop gains recomputed for each case, using the new noise propagation estimate (from step ii)
\end{enumerate}

\begin{figure}
    \begin{subfigure}[t]{1\linewidth}
    \centering
    \includegraphics[width=1\linewidth]{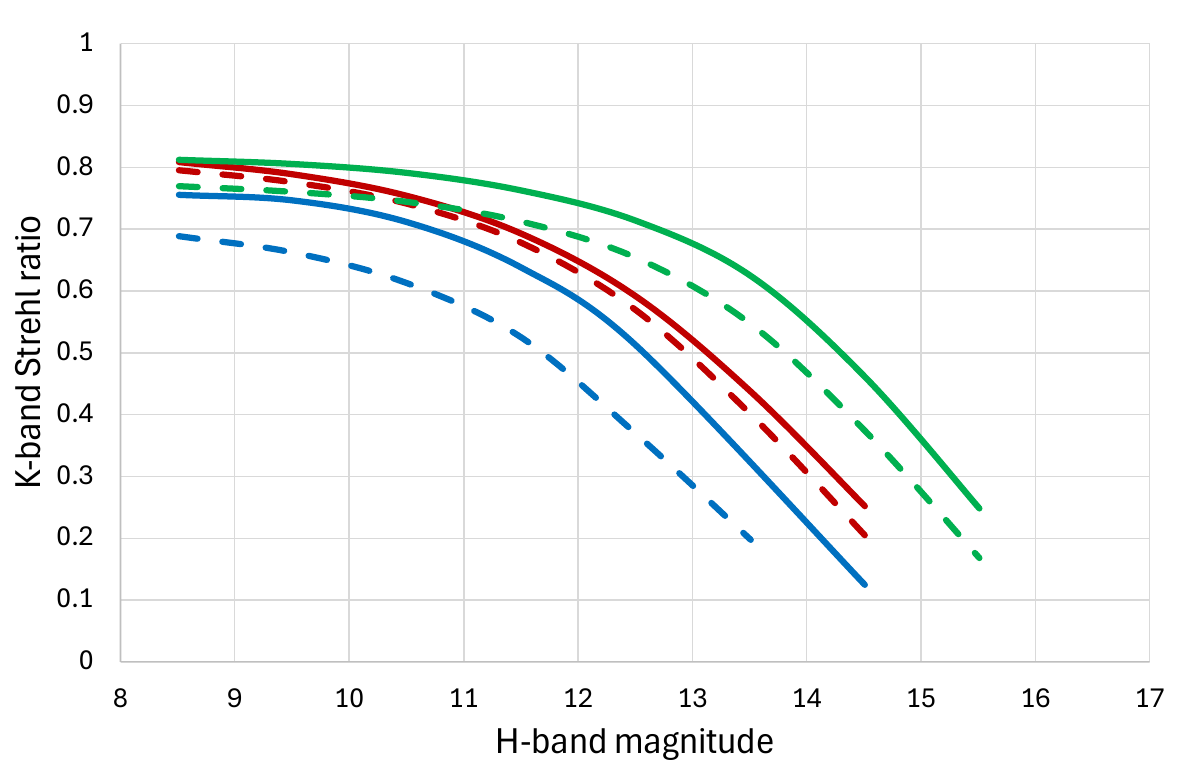}
    \caption{$r_0=8$\,cm.}
    \end{subfigure}
    \begin{subfigure}[t]{1\linewidth}
    \centering
    \includegraphics[width=1\linewidth]{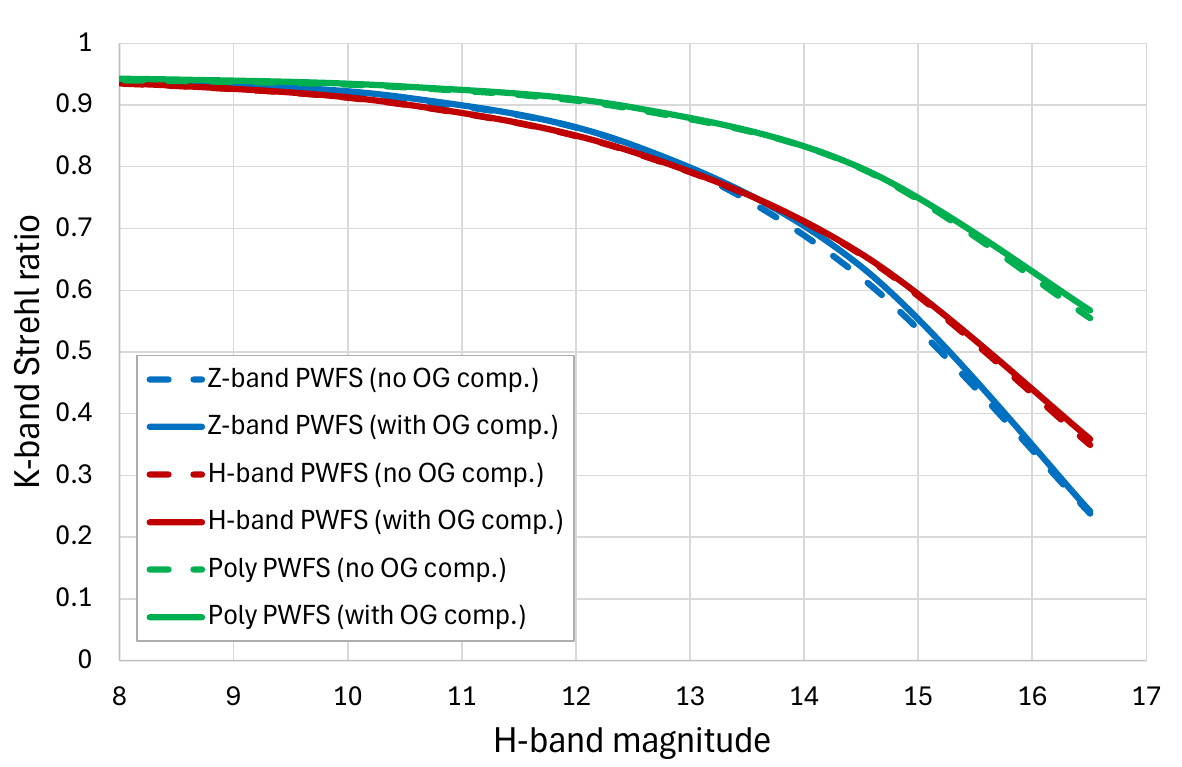}
    \caption{$r_0=16$\,cm.}
    \end{subfigure}
    \caption{Simulated AO performance (K-band Strehl ratio) versus H-band magnitude for a G-star under different atmospheric conditions, for different PWFS, with and without optical gain compensation. The results demonstrate an increase in limiting magnitude when using a polychormatic PWFS.}
    \label{fig:SRvH}
\end{figure}

Fig.~\ref{fig:SRvH} shows the end-to-end simulated performance of different PWFS vs H-band magnitude for the case of a G-star with $r_0=8$\,cm and $r_0=16$\,cm. Comparing the optical gain compensated curves, we observe that the use of the polychromatic PWFS extends the limiting magnitude by at least 1 magnitude when compared to the H-band case, and almost 2 magnitudes when compared to the $r_0=8$\,cm Z-band case. This is in agreement with Fig.~\ref{fig:IWA Strehl} where only the increase in the total number of photons was accounted for. 
It should be noted that, whilst the relative behaviour shown in Fig.~\ref{fig:IWA Strehl} and Fig.~\ref{fig:SRvH} agree, the Strehl ratio and limiting magnitude values differ, as each plot includes different errors. In this section, we only consider atmospheric turbulence and photon noise and control up to 500 KL modes, leading to a greater Strehl ratio and limiting magnitude. Fig.~\ref{fig:IWA Strehl}, on the other hand, includes other wavefront error terms (primary phasing errors, non-common path errors etc.), controls 350 modes and includes read and background flux noise, leading to a lower Strehl ratio and limiting magnitude.

In Fig.~\ref{fig:SRvH}  the optical gain compensation has a significant effect on the performance in the $r_0=8$\,cm case.  The Z-band PWFS performance is limited by the non-linear behaviour of the sensor, and therefore benefits the most from compensation of the optical gain. The H-band sensor exhibits a smaller increase in performance with optical gain compensation, demonstrating the smaller optical gain effect at longer wavelengths. These results demonstrate the potential benefits of using a polychromatic PWFS with optical gain tracking - we are able to achieve the peak performance offered by longer wavelengths for bright stars whilst extending the performance to fainter stars. While we see some increase in limiting magnitude without optical gain compensation, to fully utilise the additional photons, this compensation is required to address the greater optical gain effect at shorter wavelengths,  when operating with stronger turbulence. With increasing $r_0$, or increasing wavelength, the optical gain tracking becomes less critical, as shown for the $r_0 = 16$\,cm case in Fig.~\ref{fig:SRvH}. In this case, even as the AO residuals increase for fainter guide stars the optical gain effect mixes with the optimisation of the loop gain, resulting in similar AO performance with and without OG tracking.

It should be emphasised that the optical gain compensation for the single band sensors shown here relies on other methods to compute the optical gain, whilst the polychromatic PWFS can estimate these directly from the closed loop PWFS signals, as outlined in previous sections.

\begin{figure}
    \begin{subfigure}[t]{1\linewidth}
    \centering
    \includegraphics[width=1\linewidth]{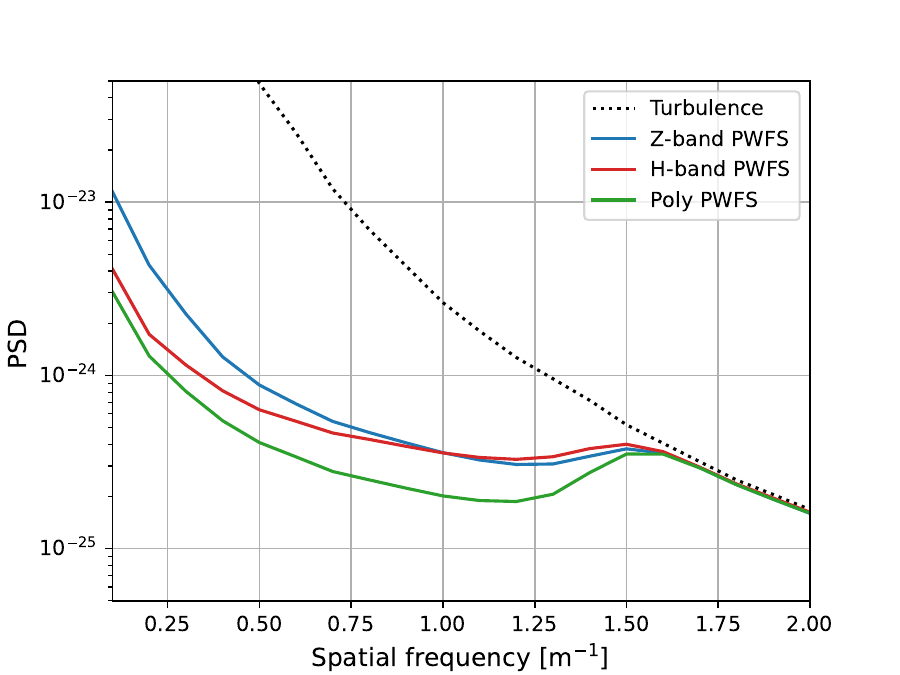}
    \caption{H-band magnitude of 10.5}
    \end{subfigure}
    \begin{subfigure}[t]{1\linewidth}
    \centering
    \includegraphics[width=1\linewidth]{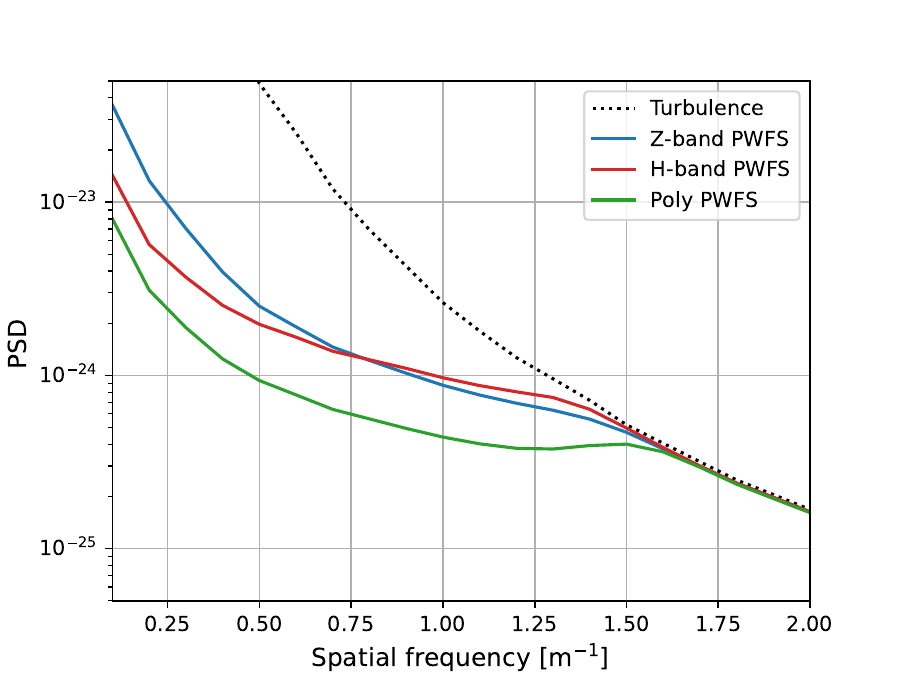}
    \caption{H-band magnitude of 12.5}
    \end{subfigure}
    \caption{Simulated radially averaged Power Spectral Density (PSD) of the closed loop residual wavefront for different PWFSs
    for a G-star with $r_0=8$\,cm.
    In each case the optical gain is compensated.
    The results demonstrate a reduction in the residual wavefront when using a polychormatic PWFS.}
    \label{fig:contrast_8cm}
\end{figure}

In addition to observing an increase in the limiting magnitude, we also observe an increase in contrast when using a polychromatic PWFS. Fig.~\ref{fig:contrast_8cm} shows the residual wavefront Power Spectral D
ensities (PSDs)  when closing the loop with different PWFS for $r_0=8$\,cm and two different H-band magnitudes (H\,$=10.5$ and H\,$=12.5$). We observe a reduction in the residual PSD within the correction band spatial frequencies ($<1.5$\,m$^{-1}$) when using a polychromatic PWFS compared to the H and Z-band PWFS. For the H\,$=10.5$ case we see an improvement of up to a factor 1.8 with the polychromatic PWFS compared to the H-band PWFS, with the greatest improvement at the mid and high spatial frequencies of the correction band. Compared to the Z-band case the polychromatic PWFS reduces the PSD across the correction band by up to a factor of 3.8, with the greatest improvement observed at the lower spatial frequencies. Similarly, for H\,$=12.5$ we observe improvements of up to 2.2 and 4.5, respectively, compared to the H-band and Z-band PWFS.

In the case of exopanet detection, the contrast improvement is important. Most planets are thought to be formed through a core accretion process, but the contrasts and separations currently achievable restrict detections to regions where the planet population density is low. The number of detectable planets is predicted to grow as instruments reach smaller angular separations and achieve deeper contrasts \citep{houlle_direct_2021}, making improvements such as the one presented here important. Moreover, improving the contrast will also improving the quality of the spectroscopic data we can obtain for these planets, which is essential to understand them.

\section{Polychromatic PWFS Practicalities}
\label{sec:practicalities}
In this section, we identify and briefly discuss practical issues associated with the design, calibration, and operation of a polychromatic PWFS. These include wavelength splitting with the science instrument, atmospheric dispersion correction, pupil registration and sampling, PWFS modulation, and NCPA. The implications on the design requirements are also discussed. An existing AO system, described in section~\ref{sec:ao_system_keck}, was selected to support evaluation of the practical issues. 

\subsection{Science Band}

A polychromatic PWFS can utilise all the photons in its wavelength band not used by the science instrument. As noted in section~\ref{sec:ao_system_keck} light must be shared between the science instrument and PWFS. We have chosen to use dichroic beamsplitters, on a motorized stage, to select the wavelength division. No physical or computational change should be needed on the PWFS side; the wavelengths not sent to the PWFS will simply not be available to be used.

\subsection{Wavefront Sensor Band}

'IWA can operate with H- or J-band light reflected by a choice of two narrow-band reflective dichroics. The polychromatic PWFS will instead operate with the entire, or a portion of the, 800 to 1800 nm band depending on the choice of dichroic. However, the AO-corrected image incident on the PWFS will vary with wavelength, with the Strehl ratio and image diameter, $\lambda/ D$,  increasing with wavelength. As the Strehl ratio drops more of the energy will go into the seeing-limited halo. The Strehl ratio variation is modest for high contrast AO systems (e.g. SPHERE-SAXO is predicted to have a $\sigma =$ 85 nm rms error for a R=5 guide star), where the polychromatic PWFS might be most beneficial, or for low order AO systems (e.g. ground-layer AO) where the Strehl ratio is low at all wavelengths.

Another factor to be considered for polychromatic PWFS usage is the colour of the guide star (see Fig. \ref{fig:Flux}). Most stars are red and M-type stars are most likely to be used for detecting and characterizing exoplanets with HISPEC and SCALES; for these stars there will be considerably more light at the red end of the 800 to 1800 nm band. Residual chromatic aberrations however can be important for high contrast systems and a polychromatic PWFS could be ideal for measuring these chromatic errors.  Dust obscured targets such as the Galactic Centre have very little flux below H-band and a 800 to 1800 nm PWFS may not offer any advantage for this target (IRS7, a red supergiant at the Galactic Centre used for wavefront sensing, would have 43 times more flux in H-band than in J-band). 

From Fig. \ref{fig:Flux}, assuming 4 pixels/subaperture and a system throughput of 0.4, a 10th magnitude K or M star would provide 20k photons/pixel/s across the Z to H-bands. A neutral density filter would be needed for stars brighter than $\sim$ 12th magnitude; MEC has a maximum count rate, to allow recovery after absorbing a photon, of 5k photons/pixel/s \citep{walter_mec_2019}. 

Sky background should be considered, especially for operation during moonlit nights where the sky will be brighter at shorter wavelengths. If necessary, wavebands with high sky contribution could be rejected or filters could be used to restrict sky contribution.

\subsection{Atmospheric Dispersion Correction}
\label{sec:ADC}
A polychromatic wavefront sensor will require atmospheric dispersion correction. 
One index of atmospheric refraction model \citep{smith_infrared_1993} is
\begin{equation}
    (n-1) 10^6 = [237.2 + 526.3 \frac{\nu_1^2}{ \nu_1^2-\nu^2}+11.69\frac{\nu_2^2}{\nu_2^2-\nu^2} ]  \frac{P_{dry}} {T}  
\end{equation}
where n is the index of refraction, $\nu$ is the wavenumber, $\nu_1$ = 114,000~cm$^{-1}$, $\nu_2$ = 62,400~cm$^{-1}$, P$_{dry}$ [kPa] is dry-air pressure, and T [K] is temperature. For a telescope on Maunakea, P$_{dry}$ = 60.5 kPa and T = 275.5 K, leading to 0.40" of dispersion between R (640 nm) and H-band (1650 nm) at 45$^{\circ}$ zenith angle. Note that alternative index of refraction models could be considered \citep{roe_implications_2002, noel_analyzing_2024} that might better model the atmosphere. 

For the reference Keck system, HISPEC and ORKID already have their own atmospheric dispersion correctors (ADCs). The Keck polychromatic PWFS can therefore have a separate ADC in its path, covering the 800 to 1800 nm wavelength band. The ADC must pass a sufficiently large field of view for acquisition; ~3 arcsec diameter should be adequate. The residual dispersion after ADC correction, and the image motion as the ADC rotates, should be small compared to the Keck diffraction limit at the shortest PWFS wavelength ($\lambda$/D = 17 mas) at zenith angles < 60$^{\circ}$. Any pupil shift or magnification introduced by the ADC should be small compared to a PWFS detector pixel (i.e. small compared to 20~cm on the Keck primary). Throughput should be high (> 90$\%$).

An ADC design that is close to meeting these requirement is the planned Gemini North AO (GNAO) system ADC \citep{rakich_personal_2025, jouve_aob_2024}. The GNAO ADC is designed for a 2 arcmin field, a 450 to 2500 nm wavelength band and a f/16.2 diverging beam. It uses two counter-rotating cemented triplet prisms and achieves < 7 mas residuals across any of the science bands (Y to K-band) for zenith angles from 0 to 60$^{\circ}$. The GNAO ADC provides an excellent starting point to design an ADC for the polychromatic PWFS's smaller field, reduced wavelength band and slightly faster beam (f/15).  

Atmospheric dispersion models are unlikely to be better than the current measurement accuracy of about 18 mas \citep{wehbe_-sky_2021} which is large compared to ELT, or even Keck, AO-corrected image size. Residual dispersion will be seen as wavelength dependent tilt by the polychromatic PWFS. This information can be used to provide slow ADC rotation control feedback \citep
{twitchell_improving_2024}. Image motion as the ADC rotates can be calibrated and used to provide slow modulator control.  

A polychromatic PWFS could potentially measure other wavelength dependent aberrations (\cite{hardy_adaptive_1998}; section 9.3.3). These are generally small but could be important for high contrast observations. If the chromatic dependent path-length errors were measured then corrections should be applied to the deformable mirror for the science wavelength. Wavelength dependent angular dispersion and multispectral errors could be minimized by sensing close to the science wavelength. 

\subsection{Pupil Registration}

If the four PWFS pupil images are not separated by an integer number of pixels, the combination of information from slightly different locations in the pupil can cause a drop in sensitivity to high spatial frequencies in the wavefront. This effect can be mitigated by oversampling of the pupil \citep{bond_adaptive_2020}, at no performance cost due to the MKIDS zero noise, as well as using intensities, instead of slopes, to directly calculate the wavefront.

Maintaining a constant pupil magnification and registration over the full 800 to 1800 nm wavelength range could prove challenging. A wavelength dependent registration could however be turned to advantage as discussed in section \ref{sec:PolyPWFS}.

\subsection{Modulation}

A polychromatic PWFS operating with a fixed modulation radius has a smaller modulating effect for longer wavelengths (by a factor of two between Z- and H-bands). 
For the lowest spatial modes the sensitivity and linear range will be dominated by the modulation, whilst for the higher order modes this is determined by the wavelength. The optimal polychromatic PWFS modulation radius versus conditions will need to be assessed.

\subsection{Non-Common Path Aberrations}

Uncorrected non-common path aberrations (NCPA) between the science and PWFS paths will impact PWFS performance. This impact was considered in the extension of the convolutional model \citep{chambouleyron_pyramid_2020} to include NCPA \citep{striffling_using_2024}. 

The Keck AO systems are calibrated during the day using a single mode fiber source (i.e. a diffraction-limited source with an optical gain of 1) at the telescope output focal plane (just inside the Rotator in Fig. \ref{fig:AOB}) that transmits 800 to 1800 nm light. The calibration process includes optimizing the image quality on the science instrument using the AO deformable mirror and subsequently determining the PWFS offsets required to address the NCPA seen on the PWFS.  

A result of using a diffraction-limited calibration source is that the wavefront can be significantly underestimated during operation and the correction under-applied (hence the on-sky gain calibration), whereas any NCPA correction is over-applied. 

The PWFS system design should therefore endeavour to minimize NCPA. The largest NCPA source is likely to be the two dichroic beamsplitters in the path to the science instrument and PWFS. These dichroics will be wedged to minimize lateral chromatic aberrations in the science path and a compensating beamsplitter can be installed in the PWFS path. Ideally a MEMS (micro-electro-mechanical system) deformable mirror would be located in the PWFS path to minimize static NCPA. 

For the polychromatic PWFS the residual NCPA offsets could be a function of wavelength due to chromatic optical aberrations introduced by the PWFS. The polychromatic PWFS could potentially provide on-sky information to compensate for NCPA errors. This effect will be included in future simulations.

\subsection{Daytime Calibrations}
A polychromatic PWFS will require additional wavelength dependent calibrations (beyond the NCPA calibrations previously discussed). Calibrations that should be checked for any wavelength dependence include pupil registration to the deformable mirror and optical gains.  In the Keck system these calibrations are performed with a single mode fiber at the input focal plane of the AO system. 

\section{Conclusion}
A method for polychromatic wavefront reconstruction in pyramid wavefront sensors has been introduced, enabled by the energy sensitivity capabilities of MKID. This approach allows simultaneous wavefront sensing over a broad spectral range, significantly enhancing the information available for wavefront reconstruction.

We have demonstrated that it is possible to estimate and compensate for optical gain variations in real time by exploiting the wavelength sensitivity of MKIDs, without the need for additional probe signals or hardware. Simulation results show that the polychromatic optical gain tracking method substantially improves the stability and performance of the AO system, particularly in poor atmospheric conditions and at lower frame rates. We show that this method is valid for a range of polychromatic configurations, achieving the best performance when employing more than two colour bands. We also demonstrate that the method can be adapted to account for noise by taking a total least squares approach, where a theoretical computation of the ratio of the expected photon noise in each colour with respect to the reference wavelength are used to un-bias the resulting optical gain ratios. These optical gain ratios can be further improved with temporal averaging (Fig. \ref{fig:ratios_TLSQ_methods}).

A series of end-to-end simulations of a polychromatic PWFS, based on the Keck II AO system, were performed. The results showed that the optical gain compensation of the polychromatic PWFS would facilitate K-band Strehl ratio increases of 0.1 at 500 Hz frame rates for a bright star in poor ($r_0$ = 8 cm) seeing, compared to the polychromatic PWFS without optical gain compensation (Fig. \ref{fig:strehl_vs_cases} and Fig.~\ref{fig:SRvH}). We also demonstrated that the polychromatic PWFS proposed for the Keck II AO application could allow $\sim$ 1 magnitude fainter guide stars to be used (compared to an H-band PWFS) and $\sim 2$ magnitudes (compared to a Z-band PWFS) depending on stellar type, due to the increased flux (Fig,~\ref{fig:IWA Strehl} and Fig.~\ref{fig:SRvH}). This benefit is only fully realised with the optical gain compensation approach shown in this paper. Finally, contrast improvements of between 1.8 and 3.8 were observed for a H = 10.5 magnitude, G-star (2.2 and 4.5 for a H = 12.5 G-star) and $r_0=8$\,cm with a polychromatic PWFS, when compared to single band PWFSs (H or Z-band; Fig.~\ref{fig:contrast_8cm}).  This contrast improvement is essential to improve the scientific understanding of exoplanets.

An overview of expected performance improvement has been provided using the Keck II AO system as a reference design, with consideration given to design practicalities. Practical considerations, including spectral band selection, calibration strategies, and system integration have been considered. These findings pave the way for the deployment of polychromatic PWFS systems on-sky, offering a promising path toward improved wavefront sensing for next-generation astronomical instruments.

As a next step, the end-to-end simulations will be extended to incorporate more realistic effects (e.g. additional noise sources and non-common path aberrations). This will allow assessment of the potential for improved sensitivity to fainter targets and to evaluate the robustness of the optical gain tracking method. In parallel, a bench demonstration will be carried out to validate the polychromatic reconstruction approach through experimental validation \citep{magniez_polychromatic_2024}. This will include the testing of MKIDs as wavefront sensor detectors, the verification of the optical gain tracking technique, and the exploration of additional methods such as super-resolution. 

\section*{Acknowledgements}
We thank Alexis Carlotti for a useful input on the contrast improvement impact on  exoplanet detection.  This project has been funded by Science and Technology Facilities Council (STFC) grants (ST/S001409/1, ST/001360/1) and STFC's Centre for Instrumentation (CfI) program. The 'IWA project is funded by the US National Science Foundation Major Research Instrumentation Program award AST-2510577.

\section*{Data Availability}

The data generated specifically for this article will be shared on reasonable request to the corresponding author.

\bibliographystyle{rasti}
\bibliography{references}
\bsp	
\label{lastpage}
\end{document}